\documentclass[11pt,a4paper]{article}
\usepackage{jheppub}
\usepackage{graphicx}
\usepackage{lipsum}
\usepackage{slashed}
\usepackage{mathtools}
\usepackage{bm}
\usepackage[all]{hypcap}
\usepackage{tikz}
\usepackage{tikz-feynman}
\newcommand{\bea}{\begin{eqnarray}}
\newcommand{\eea}{\end{eqnarray}}
\usepackage{bbm}
\usepackage{xcolor}
\usepackage{subcaption}
\usepackage{slashed}
\usepackage{booktabs}
\usepackage{multirow}
\usepackage[font=small,labelfont=bf]{caption}

\def\stw{s_{\theta_W}}
\def\ctw{c_{\theta_W}}
\def\ttw{t_{\theta_W}}
\newcommand{\eq}[1]{Eq.~(\ref{#1})}

\begin{document}
\begin{flushright}
CERN-TH-2025-144 \\
TIFR/TH/25-16
\end{flushright}
\title{An EFT study of the $pp \to \bar{t} t Z(ll) h(bb)$ process at the FCC-$\boldsymbol{hh}$}

\author[a,b]{Shankha Banerjee,}
\author[c]{Rick S. Gupta,}
\author[d]{Shilpi Jain,}
\author[e]{Michelangelo Mangano,}
\author[f]{Elena Venturini}

\affiliation[a]{The Institute of Mathematical Sciences, Taramani, 600113 Chennai, India}

\affiliation[b]{Homi Bhabha National Institute, Training School Complex, Anushakti Nagar, Mumbai 400094, India}

\affiliation[c]{Department of Theoretical Physics, Tata Institute of Fundamental Research,
Homi Bhabha Rd, Mumbai 400005, India}

\affiliation[d]{Department of High Energy Physics, Tata Institute of Fundamental Research, Homi Bhabha Rd, Mumbai 400005, India}

\affiliation[e]{CERN, Theoretical Physics Department, CH-1211 Geneva 23, Switzerland}

\affiliation[f]{Physik Department, TU München, James-Franck-Straße, 85748 Garching, Germany}
\emailAdd{shankhab@imsc.res.in}
\emailAdd{rsgupta@theory.tifr.res.in}
\emailAdd{shilpi.jain@cern.ch}
\emailAdd{michelangelo.mangano@cern.ch}
\emailAdd{elena.venturini5891@gmail.com}

\abstract{We carry out an Effective Field Theory (EFT) study of the $pp \to \bar{t} t Zh$ process in the $4b + 3 \ell + \ge 2j + \slashed{E}_T$ final state. This process can uniquely probe the $\bar{t} t Zh$ couplings arising from higher dimensional EFT operators and can also provide bounds on  $\bar{t} t Z$ coupling deviations. We highlight the importance of  the  proposed proton-proton Future Circular Collider (FCC-$hh$) to study  this process and then perform a complete collider analysis by examining the relevant background processes.  This allows us to  determine the FCC-$hh$ sensitivity to probe anomalous  $\bar{t} t Zh$ couplings.}

\maketitle

\section{Introduction}
\label{sec::Introduction}
The study of the interactions of the top quark with the electroweak and Higgs ($h$) bosons is among the most interesting targets of the physics programme of the Large Hadron Collider (LHC) and of future high-energy colliders, particularly those directly exploring the multi-TeV mass scale. These measurements enrich our understanding of the Standard Model (SM) dynamics and are a sensitive probe of possible signals of new physics beyond the Standard Model (BSM). Even after 30 years of direct measurements of the top quark, many of its interactions -- top-quark chromomagnetic and chromoelectric dipole moments~\cite{Aguilar-Saavedra:2014iga,Gaitan:2015aia, Hernandez-Juarez:2018uow, Kidonakis:2023htm},  four-fermion contact interactions~\cite{Aguilar-Saavedra:2010uur, Buckley:2015lku, DHondt:2018cww, Brivio:2019ius, Khatibi:2020mvt, DiNoi:2023ygk}, top-Higgs interactions including $CP$-violating ones~\cite{Gupta:2009wu, Englert:2014uqa, Banerjee:2019jys, Bhattacharya:2022kje, CMS:2022quh, DiNoi:2023onw, Barger:2023wbg, Bhardwaj:2023ufl}, effective right-handed charged currents~\cite{Alioli:2017ces}, top compositeness~\cite{Lillie:2007hd, Pomarol:2008bh, Darme:2021gtt} -- remain largely unconstrained experimentally or are only very weakly constrained. In particular, the $\bar{t} t Z$~\footnote{Refs.~\cite{Jiang:2025frv, Cao:2020npb} investigated constraints on the $\bar{t}tZ$ coupling at the LHC and at the proposed Electron–Ion Collider.} and $\bar{t} t Z h$~\footnote{Ref.~\cite{Maltoni:2019aot} provided preliminary Effective Field Theory sensitivity estimates for several top couplings, including $\bar{t}tZh$, by examining the energy growth of the underlying scattering amplitudes.} interactions are still poorly constrained compared to the analogous couplings of lighter quarks. The latter were precisely measured in $Z$-boson decays by LEP~\cite{Falkowski:2014tna} and will be constrained to sub-percent level precision at the HL-LHC~\cite{Banerjee:2015bla, Banerjee:2018bio, Banerjee:2019pks, Banerjee:2019twi, Araz:2020zyh, Bishara:2022vsc} and at future $e^+e^-$ machines~\cite{Amar:2014fpa, Banerjee:2021huv}. The focus of this paper is the exploration of the potential of the future $pp$ collider FCC-$hh$~\cite{FCC:2025uan} to probe the $\bar{t} t Z h$ interactions. Within the Standard Model Effective Field Theory (SMEFT) framework, our analysis would also result in projected bounds on the $\bar{t} t Z$ couplings; this is because the SMEFT  operators that generate $\bar{t} t Z h$ couplings also generate $\bar{t} t Z$ couplings. This work complements the results of Refs.~\cite{Englert:2014uqa, Banerjee:2019jys}, which showed how the FCC-$hh$ can constrain an effective $\bar{t}thh$ interaction and associated effective vertices.

The study of the top-quark’s couplings to electroweak bosons is theoretically well motivated~\cite{Malkawi:1994tg, Spira:1997ce, BessidskaiaBylund:2016jvp}. In many extensions of the SM, the top quark and its BSM partners play a key role in electroweak symmetry breaking~\cite{Grojean:2004xa, Contino:2006qr,Pomarol:2008bh, Matsedonskyi:2012ym}, often through radiative contributions to the Higgs potential~\cite{Redi:2012ha, Matsedonskyi:2012ym}. Naturalness arguments suggest that these top partners should lie near the TeV scale~\cite{Matsedonskyi:2012ym, Panico:2015jxa}, making their indirect effects potentially observable. 

The paper is organised as follows. In Section~\ref{eft}, we parametrise the EFT corrections to the process under consideration. Section~\ref{sec::Analysis} details the event generation for both signal and background samples and outlines the analysis strategy. In Section~\ref{sec::ResultsDiscussions}, we present the resulting bounds on the anomalous $\bar{t}tZh$ couplings. Finally, Section~\ref{sec::Summary} provides a summary of our findings.

\section{EFT corrections to the $pp \to \bar{t} t Z (\ell^+ \ell^-) h(b\bar{b})$ process}
\label{eft}

\begin{figure}[t]
  \centering
  \includegraphics[width=0.30\textwidth]{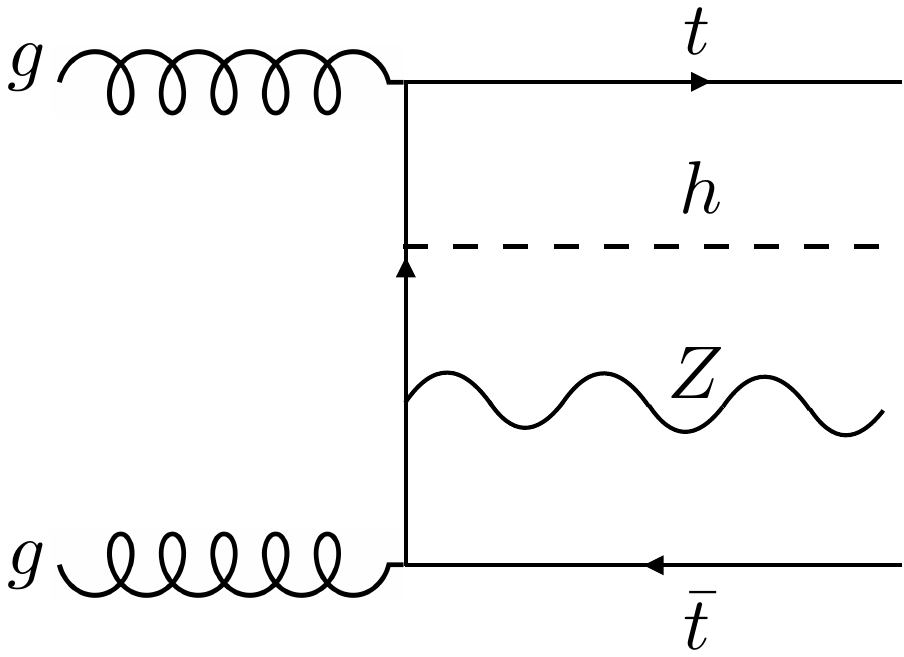}\hfill
  \includegraphics[width=0.40\textwidth]{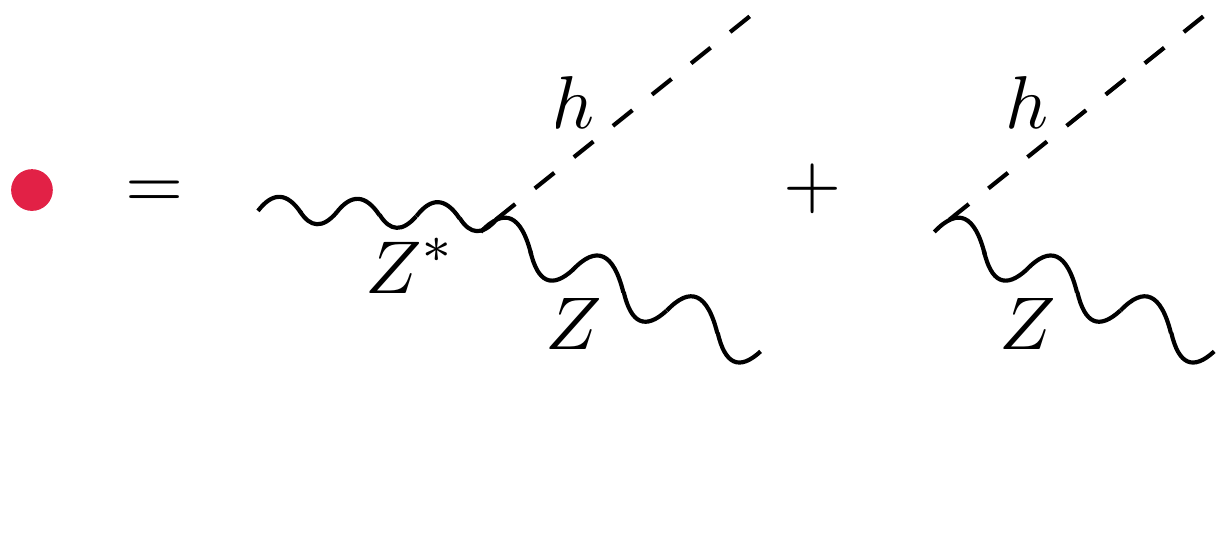}

  \vspace{0.3cm}

  \includegraphics[width=0.30\textwidth]{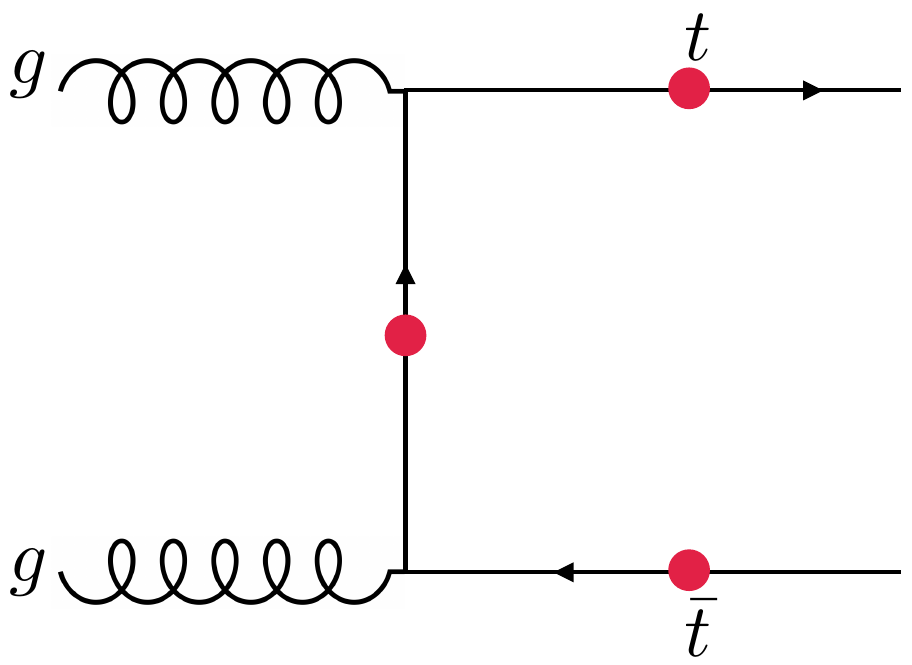}\hfill
  \includegraphics[width=0.30\textwidth]{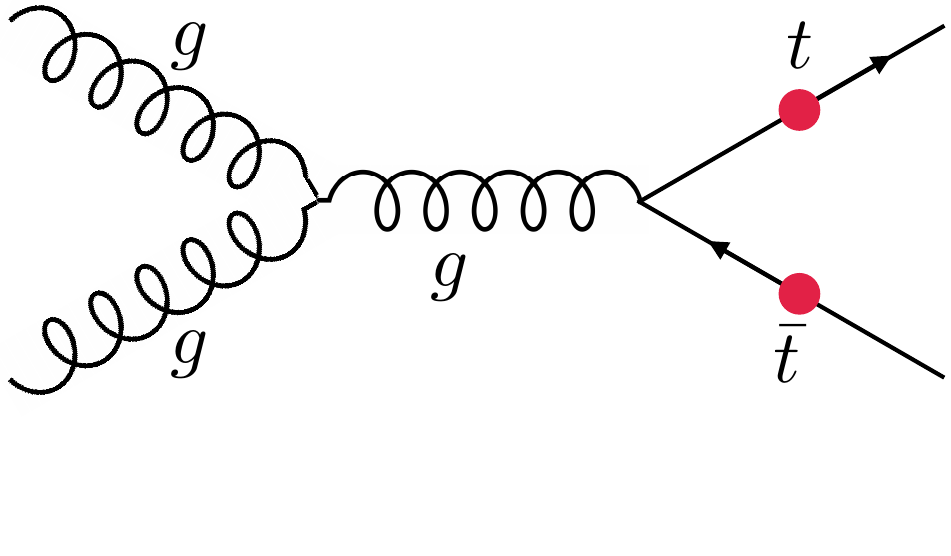}\hfill
  \includegraphics[width=0.30\textwidth]{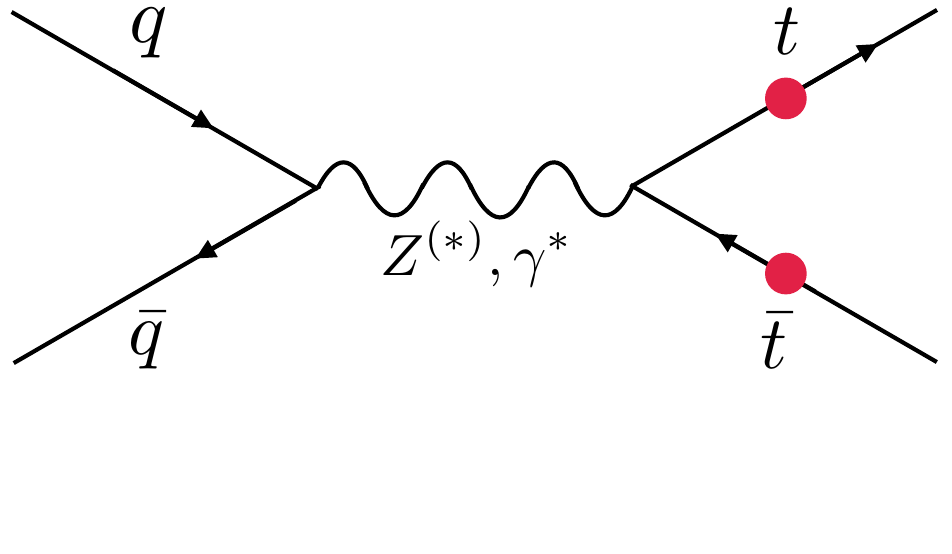}
  \caption{Representative diagrams for $gg/\bar{q}q \to \bar{t} t Zh$. 
  First: a SM-like topology with $Z$ and $h$ radiated from a top line (the emissions can occur from either of the top or anti-top lines). 
  Second: legend illustrating the symbolic red blobs used in the following panels, corresponding to either a $ZZh$ sub-diagram via a $Z$ mediator, or part of the four-point contact interaction $\bar{t}tZh$. 
  Third: $t\bar t Zh$ production with hard via a $t$-channel top exchange. 
  Fourth: $t\bar t$ production via $s$-channel $gg$ fusion. 
  Fifth: $t\bar t$ production via $s$-channel with $\bar q q$ initial states. The mediator is either a $Z$ boson or a photon. One of the coloured blobs in the last three diagrams must be replaced by the  the structures shown in the second panel. 
  The diagrams shown are representative and illustrate only a subset of the full set of contributions.}
  \label{fig:tthz_diagrams}
\end{figure}

We wish to study the $pp \to \bar{t} t Z (\ell^+ \ell^-) h(b\bar{b})$ process at FCC-$hh$ within the EFT framework. Representative Feynman diagrams contributing to $gg\to \bar{t} t Zh$ are shown in Fig.~\ref{fig:tthz_diagrams}.  A general way to parametrise corrections to this process is using anomalous couplings. The anomalous couplings up to dimension-6 level that contribute to  $pp \to \bar{t} t Z (\ell^+ \ell^-) h(b\bar{b})$ are as follows,
\bea
\Delta {\cal L}_{tthZ}&\supset& \Delta {\cal L}_{Z \ell \ell}\,+ \Delta {\cal L}_{hZZ}\,+\Delta {\cal L}_{htt}\,+\Delta {\cal L}_{hbb}\,+\Delta {\cal L}_{ggt}\,+  {\cal L}_{qqtt}\, \nonumber\\&+&  \delta g^Z_{t_L} Z_\mu \bar{t}_L \gamma^\mu t_L+  \delta g^Z_{t_R} Z_\mu \bar{t}_R \gamma^\mu t_R +g^Z_{t_{LR}} \left(\bar{t}_L \sigma^{\mu \nu} t_R Z_{\mu \nu}+h.c.\right)\nonumber\\
&+&  g^h_{Zt_L}\,\frac{h}{v}Z_\mu \bar{t}_L \gamma^\mu t_L+ g^h_{Zt_R}\,\frac{h}{v}Z_\mu \bar{t}_R \gamma^\mu t_R +\left(g^h_{Zt_{LR}} \frac{h}{v} \bar{t}_L \sigma^{\mu \nu} t_R Z_{\mu \nu}+h.c.\right).
\label{anam}
\eea
In the Higgs Effective Field Theory (HEFT), all these couplings  arise independently. In the SMEFT, on the other hand, these couplings arise in a correlated way at the dimension-6 level~\cite{Gupta:2014rxa}. In the above Lagrangian, $\Delta {\cal L}_{Z \ell \ell}$ denotes corrections to the SM $Z \ell \ell$ couplings that are already constrained very strongly by LEP data. The terms, $\Delta {\cal L}_{hZZ}\,,\Delta {\cal L}_{htt}\,$ and $\Delta {\cal L}_{hbb}$ contain anomalous couplings that respectively modify the SM $hZZ$, $htt$ and $hbb$ vertices. All these couplings can be measured at sub-percent precision by combining measurements performed at the HL-LHC and FCC-$ee$~\cite{FCC}. Finally the terms $\Delta {\cal L}_{ggt}\,$ and ${\cal L}_{qqtt}$ denote, respectively, corrections to the $tgg$ vertex and four-fermion operators with two light quarks and two top quarks. These two terms are expected to be much more strongly constrained by $t \bar{t}$ production  at the FCC-$hh$~\cite{Aguilar-Saavedra:2014iga}---a process that has a cross-section orders of magnitude higher than $pp \to\bar{t} t Z (\ell^+ \ell^-) h(b\bar{b})$ at FCC-$hh$.

\begin{table}[t]
\small
\centering
\begin{tabular}{c}
\begin{tabular}{||c||}
\hline
\rule{0pt}{4ex}
${\cal O}_{T}=\frac{1}{2}(H^\dagger \overleftrightarrow{D_\mu} H)^2$ \\
\rule{0pt}{4ex} ${\cal O}_{HWB}=  H^\dagger \sigma^a H W^a_{\mu\nu}B^{\mu\nu}$\\
\rule{0pt}{4ex}
$\mathcal{O}^{(1)}_{HQ_{L3}}=(i H^{\dagger}\overleftrightarrow{D}_{\mu}H)(\bar{Q}_L\gamma^{\mu}Q_{L3})$\\
\rule{0pt}{4ex}
    $\mathcal{O}_{HQ_{L3}}^{(3)} =(i H^{\dagger}\sigma^a\overleftrightarrow{D}_{\mu}H)(\bar{Q}_L\sigma^a\gamma^{\mu}Q_{L3})$  \\ 
    \rule{0pt}{4ex}
    $\mathcal{O}_{Ht_R} =(i H^{\dagger}\overleftrightarrow{D}_{\mu}H)(\bar{t}_R\gamma^{\mu}t_R).$
    \\
    \hline
 \end{tabular}
\end{tabular}
\caption{Dimension-6 operators in the Warsaw basis that contribute to the anomalous $Ztt$ and $hZtt$ interactions in \eq{anam}. Here  $H^\dagger {\overleftrightarrow{ D_\mu}} H\equiv H^\dagger D_\mu H - (D_\mu H)^\dagger H $, with $D_\mu H = \partial_\mu H- \frac{i \hat{g}\sigma^a}{2} W^a_\mu H - \frac{i \hat{g}'}{2} B_\mu H$ and $Q_{L3}$ is the third generation SU(2) doublet containing $b_L$ and $t_L$.}
\label{opers}
\end{table}

Next, consider the corrections to the $Ztt$ couplings in the second line of Eq.~(\ref{anam}). These couplings would be constrained at the percent level  by the 365 GeV run of the FCC-$ee$~\cite{FCC}. 

This leaves us with the $hZtt$ couplings, in the third  line of  Eq.~(\ref{anam}), that can be probed only via the $pp \to \bar{t} t Z h$ process. Among these three anomalous couplings, the dipole coupling, $g^h_{Zt_{LR}}$, is expected to be  suppressed relative to the other two  $hZtt$ couplings in the HEFT because of the extra derivative in the corresponding operator. Consider for instance the following power-counting scheme, 
\bea
\alpha_i {\cal L}_i \sim v^2 \Lambda^2 \left(\frac{h}{v}\right)^{n_{h}} \left(\frac{\partial}{\Lambda}\right)^{n_{\partial}}\left(\frac{g}{\ctw}\frac{Z_\mu}{\Lambda}\right)^{n_{Z}} \left( \frac{g_\psi\psi}{\Lambda^{3/2}}\right)^{n_\psi},
\eea
where ${\cal L}_i$ is the Lagrangian term, $\alpha_i$ is the corresponding anomalous coupling,   $\psi$ denotes the fermion fields, $n_\psi$ denotes their number, $n_h$ is the number of Higgs bosons, $n_Z$ is the number of $Z$ bosons, $n_\partial$ is the number of derivatives and $g_\psi$ represents the coupling strength of the fermions to the BSM sector being integrated out.  While $g_\psi$ can be large in certain composite Higgs models~\cite{Pomarol:2008bh, Dror:2015nkp}, we will be conservative and take $g_\psi \sim 1$ to obtain, 
\bea
g^h_{Zt_R}\sim g^h_{Zt_L}\sim \frac{g}{\ctw}\frac{v^2}{\Lambda^2},
\label{expression1}
\eea
and a further  $1/\Lambda$ suppression for the dipole coupling $g^h_{Zt_{LR}}$.  Moreover, the dipole coupling would be generated with a loop suppression in minimally coupled UV completions (as defined in Ref.~\cite{silh}). We will thus focus on the two anomalous couplings $g^h_{Zt_{L}}$ and $g^h_{Zt_{R}}$ in this work.

Note that although the Lagrangian terms corresponding to the $g^h_{Zt_{L,R}}$ and $\delta g^Z_{t_{L,R}}$ couplings are independent in the HEFT, a non-zero $g^h_{Zt_{L,R}}$ will nevertheless generate, via loop effects, a $\delta g^Z_{t_{L,R}}$ of order
\bea
\delta g^Z_{t_{L,R}}\sim \frac{y_t}{16 \pi^2}\, g^h_{Zt_{L,R}},
\eea
where $y_t$ is the top Yukawa coupling. This induces a weak indirect constraint on $g^h_{Zt_{L,R}}$. In particular, percent-level bounds on $\delta g^Z_{t_{L,R}}$ expected at FCC-$ee$ would translate into ${\cal O}(1)$ bounds on $g^h_{Zt_{L,R}}$. By contrast, our analysis leads to much stronger, percent-level bounds directly on $g^h_{Zt_{L,R}}$.

\paragraph{SMEFT parameterisation}

In  SMEFT, at the dimension-6 level, the anomalous couplings $g^h_{Zt_R}$ and $g^h_{Zt_L}$ arise in a correlated way with the $Z$ coupling deviations, $\delta g^Z_{t_L}$ and $\delta g^Z_{t_R}$. This can be seen by writing down the contributions of the Warsaw basis operators in Table~\ref{opers} to these couplings,
\bea
\label{wilson}
\delta g^Z_{f}&=&-\frac{g Y_f \stw}{\ctw^2}\frac{v^2}{\Lambda^2} c_{HWB} -\frac{g}{\ctw}\frac{v^2}{\Lambda^2}(|T_{3f}|c^{(1)}_{HQ_{L3}}-T_{3f} c^{(3)}_{HQ_{L3}}+(1/2-|T_{3f}|)c_{Ht_R})\nonumber\\&+&\frac{\delta m^2_Z}{m^2_Z}\frac{g}{2\ctw\stw^2}(T_{3f} \ctw^2+Y_f \stw^2)\nonumber\\
g^h_{Zf}&=&- \frac{2 g}{\ctw}\frac{v^2}{\Lambda^2}(|T_{3f}|c^{(1)}_{HQ_{L3}}-T_{3f} c^{(3)}_{HQ_{L3}}+(1/2-|T_{3f}|)c_{Ht_R}),\nonumber\\
\eea
where $\stw= \sin \theta_W$, $\theta_W$ being the weak mixing angle,  $f=t_L, t_R$, and $T_{3f}$ ($Y_f$) are their $SU(2)_L (U(1)_Y)$ quantum numbers. We have used $(m_W, m_Z,\alpha_{em})$ as our input parameters and the term,
\bea
\frac{\delta m^2_Z}{m^2_Z}= \frac{v^2}{\Lambda^2}\left(2 \ttw c_{HWB}-{c_{T}}\right),
\eea
arises due to the shift in the input parameter, $m_Z$. 
 
We see from \eq{wilson} that while $g^h_{Zt_R}$ and $g^h_{Zt_L}$  receive contributions only from the last three operators in Table~\ref{opers}, the $Z$ coupling deviations, $\delta g^Z_{t_L}$  and $\delta g^Z_{t_R}$, also get   contributions from the two Wilson coefficients, $c_{HWB}$ and $c_{T}$. The latter Wilson coefficients would, however, be strongly constrained by $Z$-pole and diboson observables. The contribution of these operators to $\delta g^Z_{t_L}$  and $\delta g^Z_{t_R}$---that can be constrained at the per-mille level from  the HL-LHC and FCC-$ee$ data~\cite{pomarolw, Banerjee:2018bio, grojeanww}---can therefore be safely ignored in our analysis. Ignoring these contributions we find the simple relationship, 
\bea
g^h_{Zf}= 2 \delta g^Z_{f}= - \frac{2 g}{\ctw}\frac{v^2}{\Lambda^2}(|T_{3f}|c^{(1)}_{HQ_{L3}}-T_{3f} c^{(3)}_{HQ_{L3}}+(1/2-|T_{3f}|)c_{Ht_R}).
\label{expression2}
\eea

We see that unlike the case of HEFT, the $g^h_{Zt_R}$ and $g^h_{Zt_L}$ couplings are   not independent of $\delta g^Z_{t_L}$  and $\delta g^Z_{t_R}$  in SMEFT. This means that the $Ztt$ coupling measurements at the 365 GeV run of the FCC-$ee$~\cite{FCC} would already constrain $g^h_{Zt_R}$ and $g^h_{Zt_L}$. As we will see, however, the final bounds from the $pp \to \bar{t} t Z (\ell^+ \ell^-) h(b\bar{b})$  process would be of a similar order and would add valuable complementary information even within the SMEFT framework. 

We can estimate the cut-off for a given value of $ g^h_{Zf}$ to be, 
\bea
\Lambda\sim \frac{v}{\sqrt{\ctw g^h_{Zf}/g}}.
\label{cutoff}
\eea
This estimate is valid both in the HEFT (see \eq{expression1}) and the SMEFT (see \eq{expression2}), where we take the Wilson coefficients to be ${\cal O}(1)$ in the latter case. To ensure EFT validity, while excluding a particular value of the coupling, $g^h_{Zf}$, we will not consider events beyond the cut-off given by \eq{cutoff}.

In Fig.~\ref{fig:tthz_diagrams} we show a representative set of Feynman diagrams for our process. For the EFT contribution we display only the corrections induced by the $hZff$ contact terms, shown in the second row. The SM contribution from these diagrams is obtained by radiating a $Zh$ system via a virtual $Z$ boson from one of the red blobs. The EFT contribution instead arises from replacing one of the red blobs by the $hZff$ contact interaction. The latter differs from the SM contribution due to the absence of the $Z$ propagator. As a consequence, the EFT contribution grows quadratically with the $Zh$ invariant mass relative to the SM contribution. We discuss this behaviour in more detail in Appendix~\ref{amplitude}.

We note that electroweak radiative corrections are expected to induce Sudakov logarithmic suppressions in the multi-TeV regime. At energies $\sqrt{\hat s}\gg m_W$, virtual EW corrections to amplitudes are dominated by double-logarithmic terms of the form
\(
\delta\mathcal M/\mathcal M \sim
- \frac{\alpha}{4\pi s_W^2}\, C \log^2(\hat s/m_W^2)
\),
where $C=\mathcal O(1)$ encodes the electroweak charge factors of the external states~\cite{Ciafaloni:2000df,Denner:2000jv,Denner:2001gw,Melles:2001dh}. For $m_{Zh}\sim 1$--$1.5$~TeV this corresponds to an amplitude-level suppression of order $5$--$15\%$, or $10$--$30\%$ at the cross-section level. Such effects predominantly lead to an overall suppression of the high-energy tails of distributions, as seen in explicit NLO EW studies of multi-TeV processes at hadron colliders~\cite{Denner:2015fca,Kallweit:2015dum}. Moreover, approximate Sudakov reweighting in a SMEFT context has been studied in Ref.~\cite{Banerjee:2024eyo}. Since both the SM and EFT contributions originate from the same external electroweak charges, they are expected to receive Sudakov corrections of similar parametric form. We note that the magnitude of the effects estimated above is comparable to the typical increase in cross sections due to NLO QCD corrections. The qualitative features underlying our analysis strategy, in particular the enhanced sensitivity to $\bar{t}tZh$ contact interactions in the large-$m_{Zh}$ region, therefore remain unchanged.

For electroweak corrections in the EFT scenario, the Sudakov charge factors themselves do not depend on the EFT couplings, but the EFT effects enter through the hard amplitude $\mathcal M$, and may also appear in subleading (single-logarithmic) terms. A full NLO EW calculation including EFT effects is beyond the scope of this exploratory study. We thus expect our qualitative conclusions to remain unchanged.


Finally, before moving to the next section let us comment on the the $pp \to \bar{t}t h j$ process at the FCC-$hh$ that can also probe $hZtt$ anomalous couplings  as discussed in Ref.~\cite{Dror:2015nkp}; the authors proposed that the quadratic growth of the $tZ \to th$ amplitude with respect to the SM can be utilised to put bounds on the $hZtt$ contact terms. In this case, however, the background from gluon-initiated $pp \to \bar{t}t h j$ production is large and a preliminary analysis in Ref.~\cite{Brooijmans:2020yij} indicated that the bounds obtained from this process may not be larger than that obtained from the $pp \to \bar{t}t Z$ process. Nevertheless, a detailed collider study of this process at the FCC-$hh$ would provide a useful and complementary probe of the $hZtt$ anomalous couplings.

\section{Analysis Strategy}
\label{sec::Analysis}
Before proceeding with the analysis, we detail our event generation methodology. For the $\bar{t}thZ$ samples, we encode the SMEFT vertices in the broken phase using the \texttt{FeynRules}~\cite{Alloul:2013bka} package from which we obtain the Universal FeynRules Output (\texttt{UFO})~\cite{Degrande:2011ua} model files. We choose \texttt{Sherpa} version 2.2.11~\cite{Sherpa:2019gpd} to generate our event samples and employ \texttt{Comix}~\cite{Gleisberg:2008fv} as our tree-level matrix element generator. All of our samples are generated, for a $pp$ collider at 100 TeV, at the leading order (LO) in perturbation theory, in view of the complexity of the processes under consideration. For the parton distribution function, we choose the \texttt{CT14nlo} set within the \texttt{LHAPDF6}~\cite{Buckley:2014ana} framework. Considering that the whole study is done at the LO, the impact of choosing alternative PDF sets is marginal. For several of the backgrounds (see Tab.~\ref{tab::generation}), we use the CKKW~\cite{Catani:2001cc} ME-PS merging scheme with a QCUT value of 20 GeV. We merge these samples either with two additional jets ($\bar{t}th, \bar{t}tZ, W^+W^-Z, W^{\pm}Z, ZZZ$) or with one additional jet ($\bar{b}bW^{\pm}Z, \bar{t}tW^{\pm}Z$). We do not impose any other $K$-factors to our signal and background samples, and the Higgs decay branching ratios reflect the SM values.

In Tab.~\ref{tab::generation}, we show the various processes generated, their generation-level cuts (if any), the cross-sections, and the factorisation, and renormalisation scales chosen. The $\bar{t}tZh$ processes have the following decay chain: $h \to \bar{b}b$, and $Z \to \ell^+ \ell^-$ $(\ell = e, \mu, \tau)$. Finally, we do not include in the table the scenario where $h \to \tau^+\tau^-$ and $Z \to \bar{q}q$. For the last case, we find that the number of events surviving after all analysis cuts is negligible. The generation of the $\bar{t}tZh$ EFT samples is divided into two phase-space regions, $m_{hZ} > 750$ GeV and $m_{hZ} > 1700$ GeV. The reason for splitting the phase space is to ensure sufficient statistics for the high-energy tails of the distribution. Eventually, we record the showered events in the \texttt{HepMC3} format~\cite{Buckley:2019xhk}.

\begin{table}[htbp]
  \centering
\begin{tabular}{|l||l|l|l|r|}
\hline
\textbf{Process} & \textbf{Cut} & $\boldsymbol{\mu_F^2}$ & $\boldsymbol{\mu_R^2}$ & $\boldsymbol{\sigma}$ \textbf{(fb)} \\
\hline
$\bar{t}tZh$ (SM) & None & $\frac{1}{4} H_T^2 + 2 m_t^2 + m_h^2 + m_Z^2$ & $\frac{1}{4} H_T^2 + 2 m_t^2$ & 6.89 \\
$\bar{t}tZh$ (SM) & $m_{hZ} > 1700$ GeV & " & " & 0.13 \\
\hline
$\bar{t}th + 0, 1, 2$ jets & None & $\frac{1}{4} H_T^2 + 2 m_t^2 + m_h^2$ & $\frac{1}{4} H_T^2 + 2 m_t^2$ & 2382.72 \\
$\bar{t}t\bar{b}bh$ & None & $\frac{1}{4} H_T^2 + 2 m_t^2 + m_h^2$ & $\frac{1}{4} H_T^2 + 2 m_t^2$ & 23.16 \\
$\bar{t}tZ + 0,1,2$ jets & None & $\frac{1}{4} H_T^2 + 2 m_t^2 + m_Z^2$ & $\frac{1}{4} H_T^2 + 2 m_t^2$ & 2998.82 \\
$\bar{t}tZZ$ & None & $\frac{1}{4} H_T^2 + 2 m_t^2 + 2 m_Z^2$ & $\frac{1}{4} H_T^2 + 2 m_t^2$ & 74.91 \\
$W^+W^-Z + 0,1,2$ jets & None & $\frac{1}{4} H_T^2 + 2 m_W^2 + m_Z^2$ & $\frac{1}{4} H_T^2$ & 345.28 \\
$W^{\pm}ZZ + 0,1,2$ jets & None & $\frac{1}{4} H_T^2 + m_W^2 + 2 m_Z^2$ & $\frac{1}{4} H_T^2$ & 240.55 \\
$ZZZ + 0,1,2$ jets & None & $\frac{1}{4} H_T^2 + 3 m_Z^2$ & $\frac{1}{4} H_T^2$ & 96.53 \\
$\bar{b}b W^{\pm} Z+ 0,1$ jet & None & $\frac{1}{4} H_T^2 + m_W^2 + m_Z^2$ & $\frac{1}{4} H_T^2$ & 73.45 \\
$\bar{t}tW^{\pm} \; \textrm{(had) } Z + 0,1$ jet & None & $\frac{1}{4} H_T^2 + 2 m_t^2 + m_W^2 + m_Z^2$ & $\frac{1}{4} H_T^2 + 2 m_t^2$ & 4.38 \\
$\bar{t}tW^{\pm} \; \textrm{(lep) }Z + 0,1$ jet & None & " & " & 3.69 \\
$\bar{t}thh$ & None & $\frac{1}{4} H_T^2 + 2 m_t^2 + 2 m_h^2$ & $\frac{1}{4} H_T^2 + 2 m_t^2$ & 49.38 \\
\hline
\end{tabular}
\caption{Cross-sections (in fb) along with the generation-level cuts applied to each process. The table also lists the factorisation and renormalisation scales chosen for the processes. The signal (EFT) cross-sections are not shown explicitly.}
  \label{tab::generation}
\end{table}

The process in question is $pp \to \bar{t}tZh$, where we consider the following final state: $4b+ 3 \ell + \ge 2j+\slashed{E}_T$, with $b, j, \ell$ being a $b$-tagged jet, a light jet, and an isolated electron or muon, respectively. For the first set of cuts which mimic the trigger, we use the following,
\begin{align}
p_{T,\ell/\gamma}^{min} &= 10 \; \textrm{GeV}, p_{T,j}^{min \; (max)} = 30 \; (15000) \; \textrm{GeV}, \nonumber \\
|\eta_{\ell/\gamma}^{max}| &= 2.5,  |\eta_{j \; (b)}^{max}| = 4.5 \; (4.0),  |\eta_{b}^{mid}| = 2.5 \nonumber
\end{align}

The jets are clustered within the \texttt{FastJet}~\cite{Cacciari:2011ma} framework using the anti-$k_t$ jet-clustering algorithm~\cite{Cacciari:2008gp} with a radius parameter $R = 0.4$. The first step in the $b$-tagging algorithm is to ensure that the $b$-hadrons lie within a cone size of $\Delta R_{b} = 0.2$ of the jet. These jets are then required to satisfy $p_T$ and $\eta$-dependent FCC-specific~\cite{Selvaggi:2717698} efficiency factors as listed in Tab.~\ref{tab::tagging-eff}. We must note that some of the largest cross-sections come from processes without 4 $b$-quarks at the matrix-element level. We thus consider all possible backgrounds at LO, which can mimic our final state, which include charm and other light jets being mistagged as $b$-jets. The mistag efficiencies are also listed in Table~\ref{tab::tagging-eff}. 

In order to estimate the number of $c$-tagged jets faking $b$-jets, the jets are first matched onto charm-hadrons within $\Delta R_{c} = 0.2$. Following this, all of these $c$-tagged jets that fail the $p_T$ and $\eta$ criteria are considered as \textit{light} jets, namely jets that are neither tagged as $b$- nor as $c$-jets. For the remainder of these $c$-tagged jets that pass the aforementioned criteria, a flat random number between 0 and 1 is considered. If this random number is less than the quoted $c$-jet mistag rate as shown in Table~\ref{tab::tagging-eff}, the jet is selected as a possible $b$-tagged jet candidate. If not, this jet is given the status of a \textit{light} jet. In an analogous manner, to estimate the contribution of light jets faking $b$-tagged jets, all the light jets, barring the ones identified with $b$- or $c$-hadrons and which failed the $p_T$, $\eta$, or efficiency criteria, are considered. Following Table~\ref{tab::tagging-eff}, the light-jet to $b$-tagged jet mistag rates are applied.

Finally, we require 4 $b$-tagged jets which satisfy all the criteria described above. Furthermore, we require 3 isolated electrons or muons. We use a simple lepton isolation criteria whereby we demand that the hadronic activity around $\Delta R = 0.3$ of an electron or a muon must be less than 10\% of its $p_T$. We also include all the light jets which are within the aforementioned $p_T$ and $\eta$ ranges. 
\begin{table}[htbp]
\centering
\renewcommand{\arraystretch}{1.3}
\begin{tabular}{llcc}
\toprule
\textbf{} & \boldmath{$p_T$} \textbf{(GeV)} & \boldmath$|\eta| < 2.5$ & \boldmath$2.5 < |\eta| < 4.0$ \\
\midrule
\multicolumn{4}{c}{\textbf{b-tagging efficiency}} \\
\midrule
\multirow{2}{*}{} & $30 < p_T < 15000$ GeV & $(1 - p_T~[\text{TeV}]/15) \cdot 85\%$ & $(1 - p_T~[\text{TeV}]/15) \cdot 64\%$  \\
                  & $p_T > 15000$ GeV & $0\%$ & $0\%$ \\
\midrule
\multicolumn{4}{c}{\textbf{c-quark mistag rate}} \\
\midrule
\multirow{2}{*}{} & $30 < p_T < 15000$ GeV & $(1 - p_T~[\text{TeV}]/15) \cdot 5\%$ & $(1 - p_T~[\text{TeV}]/15) \cdot 3\%$   \\
                  & $p_T > 15000$ GeV & $0\%$ & $0\%$\\
\midrule
\multicolumn{4}{c}{\textbf{light-jet mistag rate}} \\
\midrule
\multirow{2}{*}{} & $30 < p_T < 15000$ GeV & $(1 - p_T~[\text{TeV}]/15) \cdot 1\%$ & $(1 - p_T~[\text{TeV}]/15) \cdot 0.75\%$ \\
                  & $p_T > 15000$ GeV & $0\%$ & $0\%$\\
\bottomrule
\end{tabular}
\caption{Efficiencies for $b$-tagging and mistag rates for $c$- and light-jets, as functions of jet $p_T$ and pseudorapidity $|\eta|$. Beyond $|\eta|>4$ the efficiency drops to 0. }
\label{tab::tagging-eff}
\end{table}

\begin{figure}[htbp]
  \centering
  
  \includegraphics[width=0.45\textwidth]{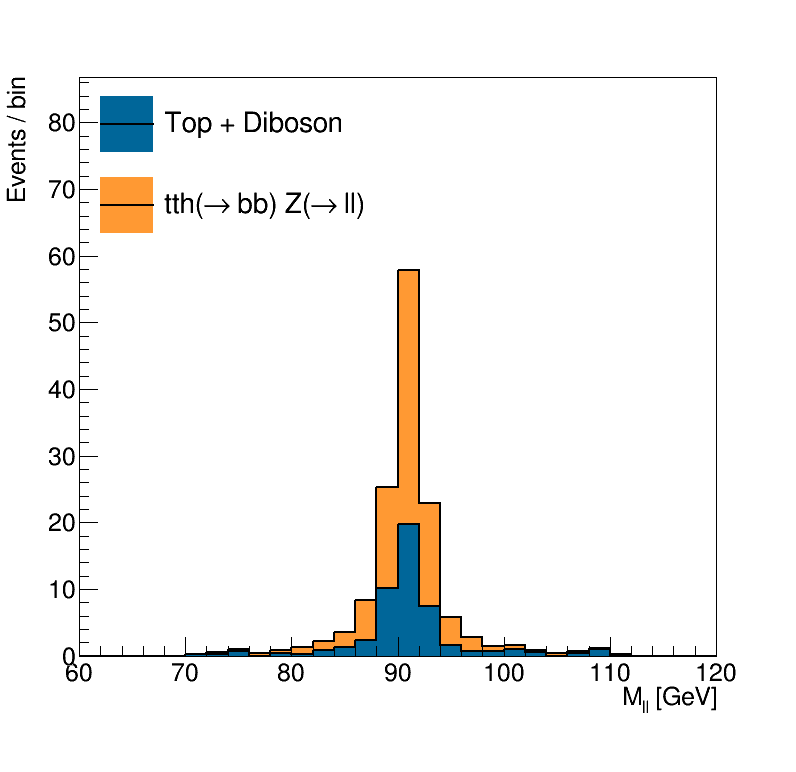}
  \hfill
  \includegraphics[width=0.45\textwidth]{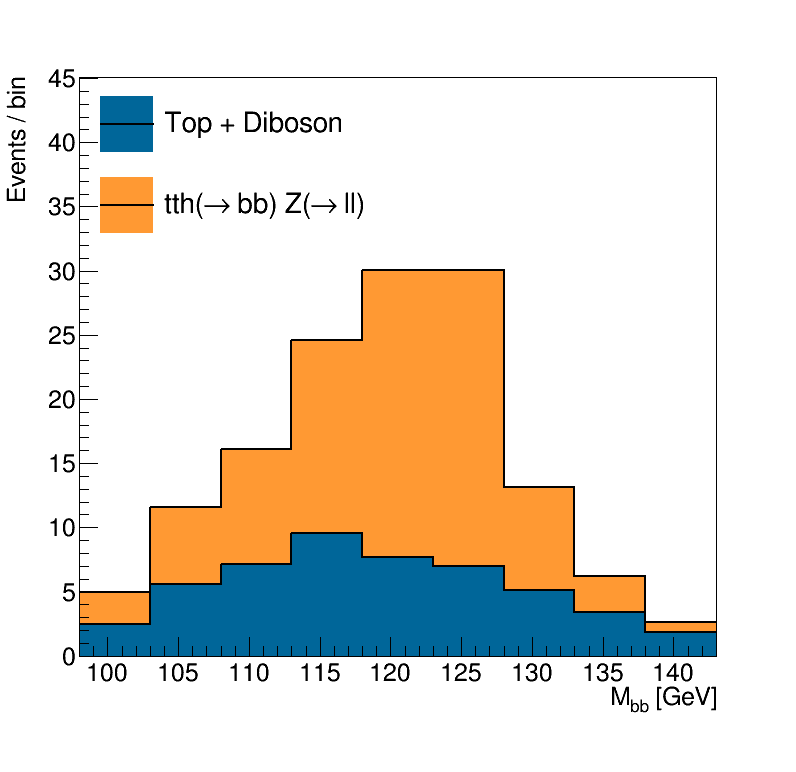}
  
  \vspace{0.5cm}
  
  \includegraphics[width=0.45\textwidth]{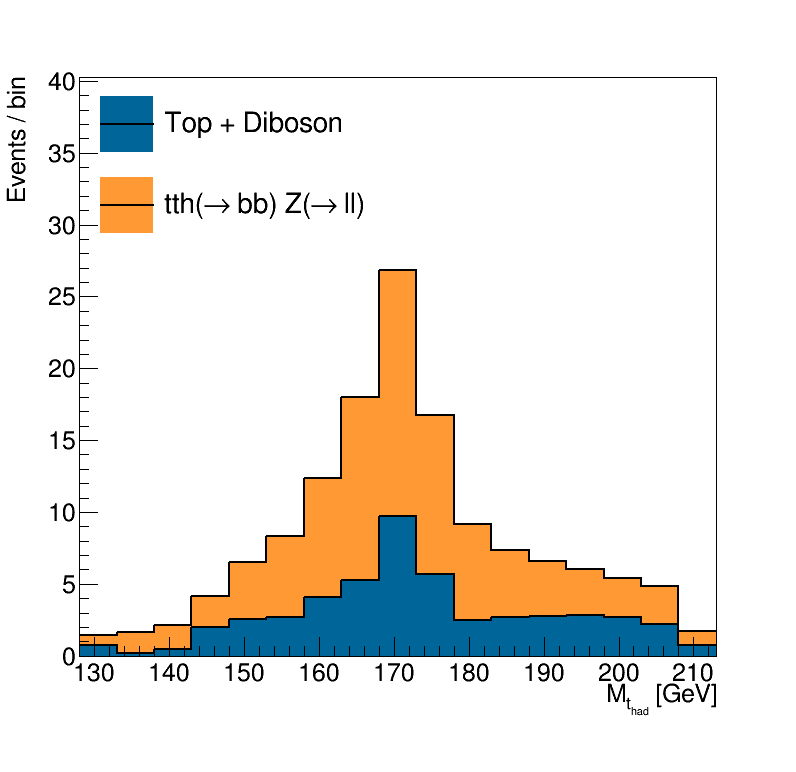}
  \hfill
  \includegraphics[width=0.45\textwidth]{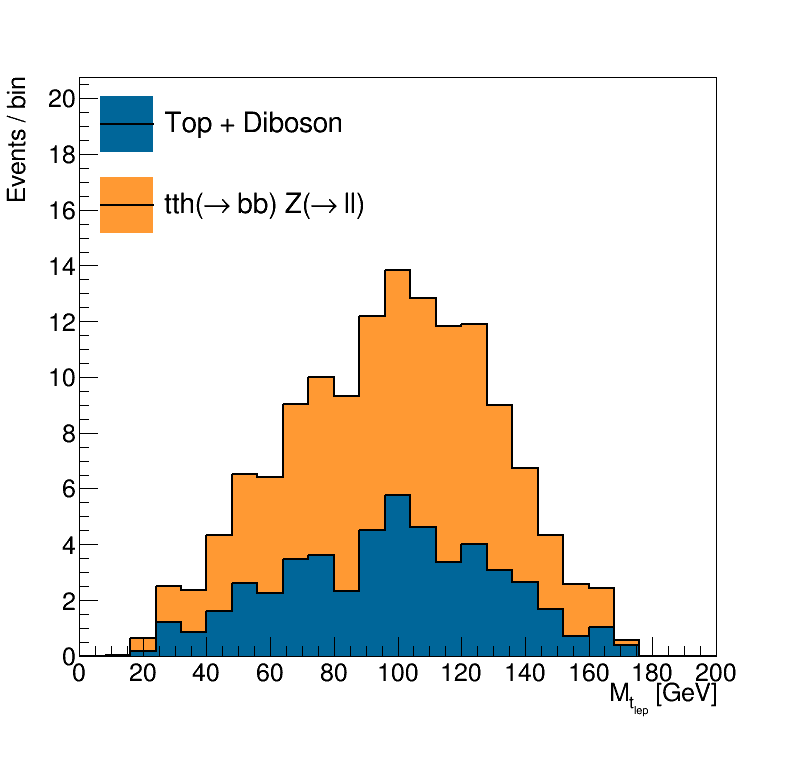}
  
  \caption{Figure shows, as stacked histograms, the reconstructed masses of the $Z$-boson, the Higgs boson, the hadronic top, and the leptonic top, respectively. The SM $\bar{t}tZh$ sample is denoted in orange and the remaining top + diboson backgrounds are represented in teal. All rates are normalised to the total integrated luminosity of 30 ab$^{-1}$.}
  \label{fig::reco_masses}
\end{figure}
\begin{figure}[htbp]
  \centering
  
  \includegraphics[width=0.45\textwidth]{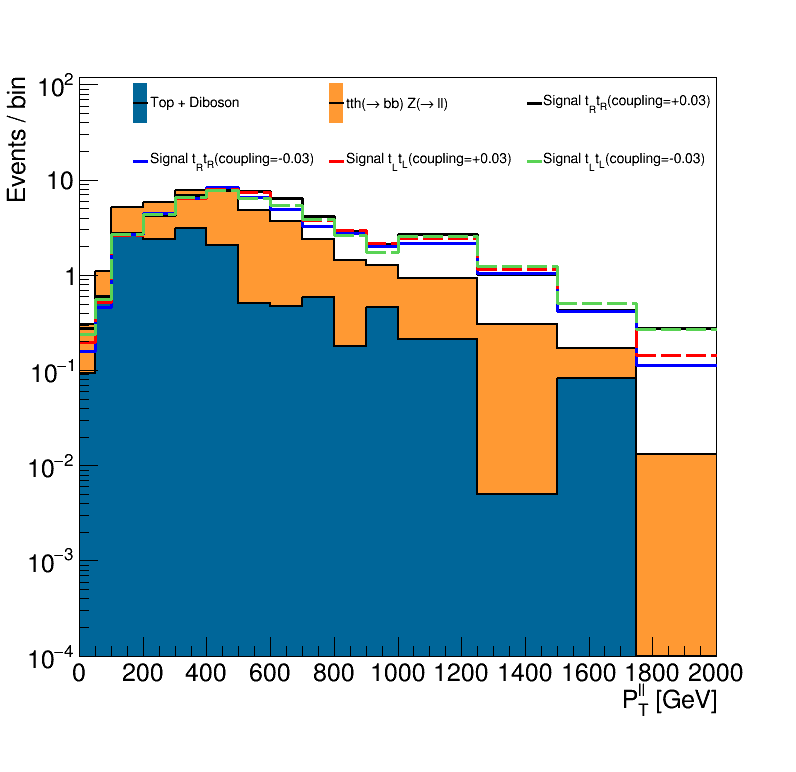}
  \hfill
  \includegraphics[width=0.45\textwidth]{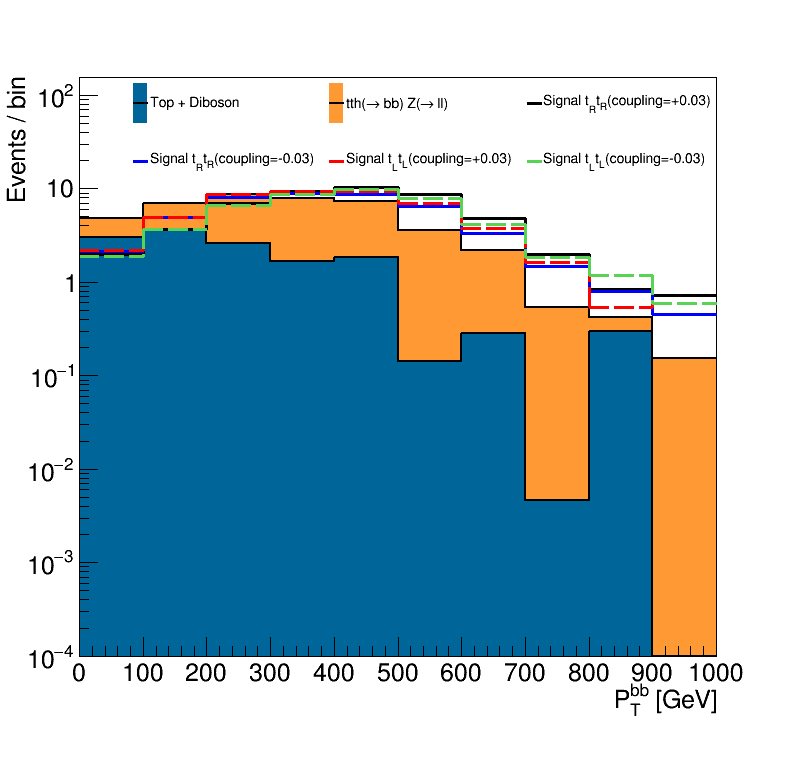}
  
  \vspace{0.5cm}
  
  \includegraphics[width=0.45\textwidth]{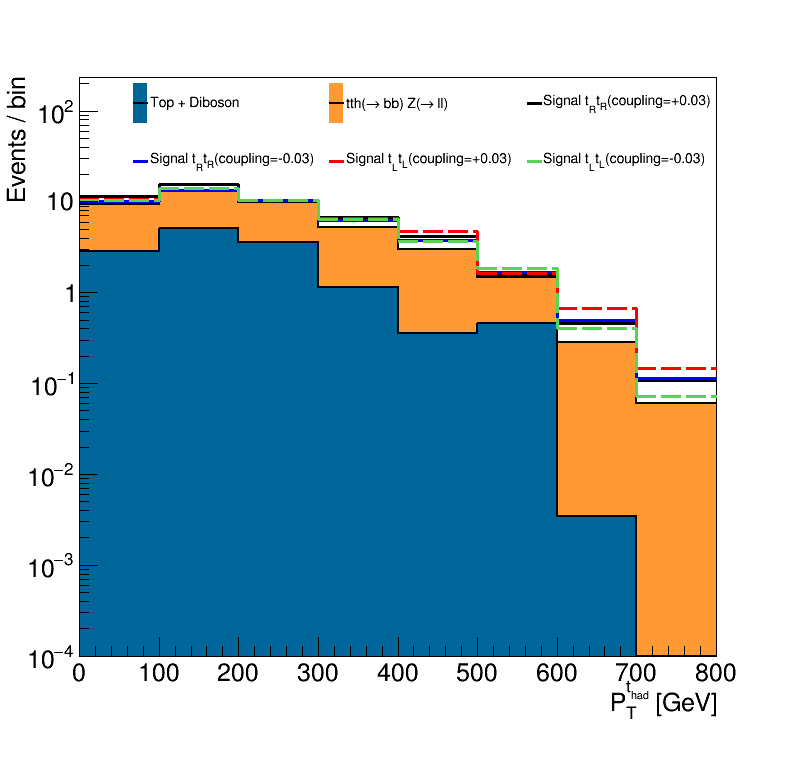}
  \hfill
  \includegraphics[width=0.45\textwidth]{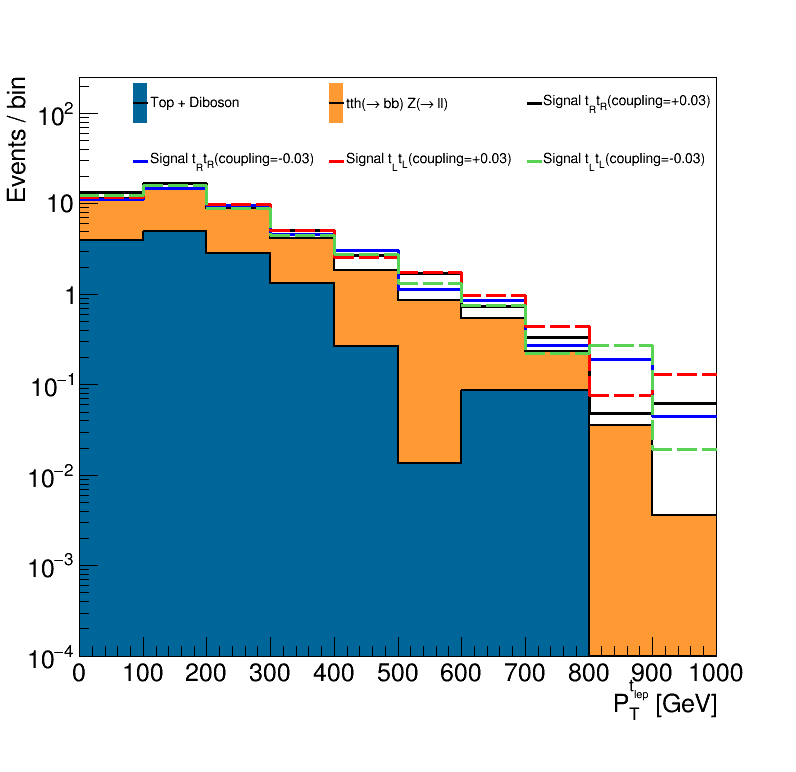}
  
\caption{Stacked histograms of the transverse momentum ($p_T$) distributions for the reconstructed $Z$ boson, the Higgs boson, the hadronic top, and the leptonic top, respectively. The plots are shown on a logarithmic scale. The colour coding for the SM background samples follows the same convention as before. Overlaid on these are the signal distributions (not stacked): the cases with couplings $g_{hZt_R} = +0.03$ ($-0.03$) are shown in black (blue), while those with $g_{hZt_L} = +0.03$ ($-0.03$) are shown in red (green).}
  \label{fig::reco_pt}
\end{figure}
\begin{figure}[htbp]
  \centering
  
  \includegraphics[width=0.45\textwidth]{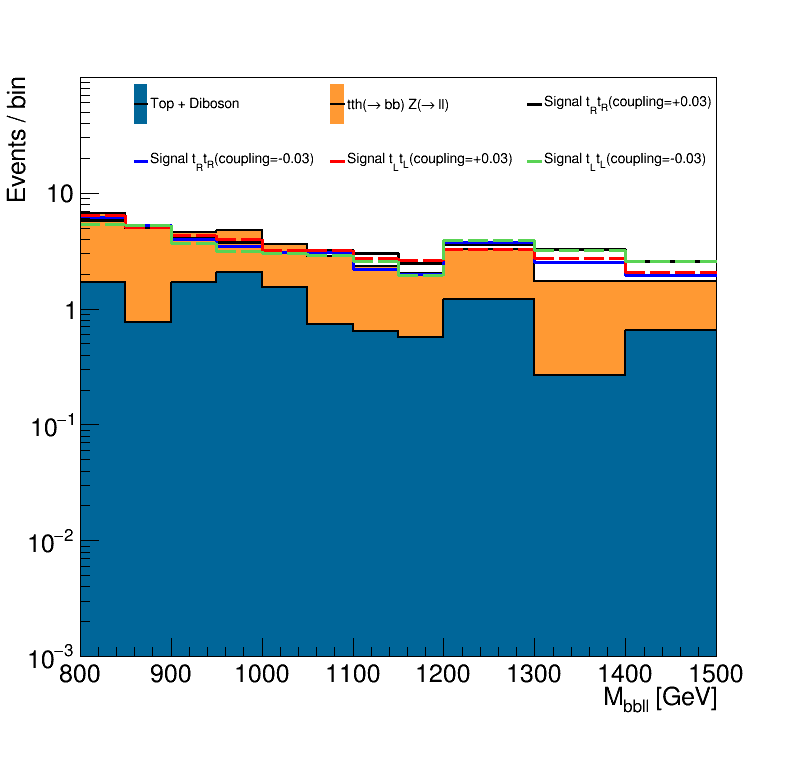}
  \hfill
  \includegraphics[width=0.45\textwidth]{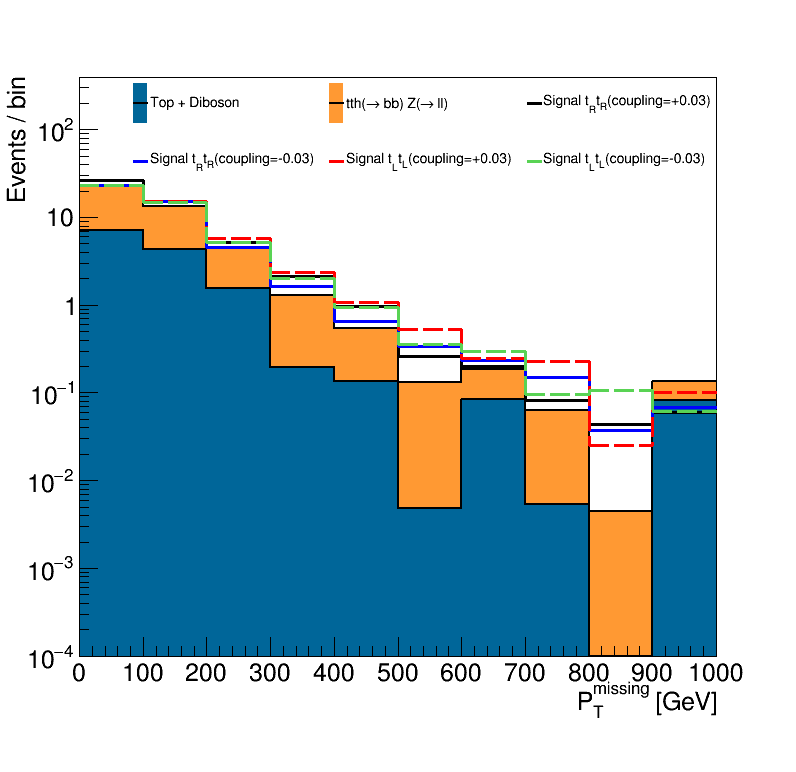}
  
\caption{Stacked histograms of $m_{Zh}$ and $\slashed{E}_T$ for the SM samples, with the signal samples overlaid (not stacked). The distributions are shown on a logarithmic scale. The labels and colour coding follow the same convention as before.}
  \label{fig::reco_misc}
\end{figure}

In order to extract the signal, we fully reconstruct the $Z$-boson, the Higgs boson, and the hadronically decaying top quark, and partially reconstruct the leptonically decaying top quark. To reconstruct the $Z$-boson, we consider the 3 charged isolated leptons. If there is a single pair of opposite-sign-same-flavour (OSSF) leptons, we reconstruct the invariant mass with them. If we have two pairs of OSSF leptons, we check which of the invariant masses is closest to $m_Z = 91.1876$ GeV. We reserve the third isolated lepton for later. Now, before we reconstruct the Higgs and the hadronic top, we require all 6 pairs of $b$-tagged jets to have invariant mass greater than 50 GeV. To then select the Higgs-decay candidates, we scan over the 6 pairs of $b$-tagged jets to minimise the following $\chi^2$:

\begin{equation}
\chi^2_h = \frac{(m_{b_ib_j}-m_h)^2}{\Delta_h^2},
\end{equation}
where $m_{b_ib_j}$ is the invariant mass formed by the pair of $b$-tagged jets $b_i$ and $b_j$, and $i \ne j$ run over all the 4 $b$-tagged jets. To perform the minimisation, we choose $m_h = 120$ GeV, and $\Delta_h = 20$ GeV as our parameters. This non-standard choice for $m_h$ requires elucidation. Given that the Higgs bosons decay to $b$-quarks, which then hadronise to $b$-hadrons, the invisible decays of the hadrons stemming from decays to neutrinos are responsible for shifting the reconstructed Higgs peak to smaller values. If we were to choose $m_h = 125$ GeV, then we would need to explicitly include jet energy correction effects in the $b$-tagged jets. Finally, after minimising $\chi^2_h$, we require that $|m_{b_ib_j}-m_h| < \Delta_h$. We consider the two remaining $b$-tagged jets and all the light jets respecting the $p_T$ and $\eta$ ranges and define our second $\chi^2$ to reconstruct the hadronic top with

\begin{equation}
\chi^2_{t_h} = \frac{(m_{b_i j_a j_b}-m_t)^2}{\Delta_t^2},
\end{equation}
where $m_{b_i j_a j_b}$ is the invariant mass formed by one of the two remaining $b$-tagged jets, and two of the light jets. $i$ runs over the remaining 2 $b$-tagged jets and $a \ne b$ denote the indices of every light jet. We require $\Delta_t = 40$ GeV as we expect the uncertainty in the reconstructed mass to be larger owing to it being reconstructed from three jets. Finally, we require $|m_{b_i j_a j_b}-m_t| < \Delta_t$ GeV. Finally, we partially reconstruct the leptonic top quark with the remaining $b$-tagged jet and the isolated lepton which didn't reconstruct the $Z$-boson. The reconstructed $Z$, Higgs, and top quarks in the SM are shown in Fig.~\ref{fig::reco_masses}, with the SM contributions displayed as stacked histograms. For clarity, EFT signal samples are omitted from these plots. Figure~\ref{fig::reco_pt} presents the transverse momentum distributions of the same four reconstructed objects, shown on a logarithmic scale, with the SM still displayed as stacked histograms. In addition, four signal curves corresponding to $g_{hZt_{R/L}} = \pm 0.03$ are overlaid. Finally, Fig.~\ref{fig::reco_misc} displays the invariant mass of the $Zh$ system and the missing transverse momentum. 

   
\section{Results and Discussions}
\label{sec::ResultsDiscussions}

We perform a $\chi^2$ fit in the 2D plane of $\bar{t}_{R}t_{R}hZ$ and $\bar{t}_{L}t_{L}hZ$. We define our $\chi^2$ as

\begin{equation}
 \chi^2 = \sum_i \frac{\left[ \mathcal{O}_{i}^{\text{theo.}}(g_{hZt_R}, g_{hZt_L}) - \mathcal{O}_{i}^{\text{exp.,SM}} \right]^2}{\sigma_{i}^2},
\end{equation}
where the theoretical prediction in each bin~\footnote{Following Fig.~\ref{fig::reco_misc}, we have used 11 bins in the range 800 GeV $< m_{Zh} <$ 1.5 TeV. Thus, $i$ runs from 1 to 11.} $i$ of the $m_{Zh}$ distribution is given by
\begin{align}
\mathcal{O}_i^{\text{theo.}}(g_{hZt_R}, g_{hZt_L}) & =
\mathcal{O}_i^{\text{SM}} 
+ g_{hZt_R} \mathcal{O}_i^{\text{INT}, \bar{t}_R t_R h Z} 
+ g_{hZt_L} \mathcal{O}_i^{\text{INT}, \bar{t}_L t_L h Z}
+ g_{hZt_R}^2 \mathcal{O}_i^{\text{SQ}, \bar{t}_R t_R h Z} \nonumber \\
& + g_{hZt_L}^2 \mathcal{O}_i^{\text{SQ}, \bar{t}_L t_L h Z} 
+ g_{hZt_R} g_{hZt_L} \mathcal{O}_i^{\text{MIX}}.
\label{eq:Oi_theo}
\end{align}

Here:
\begin{itemize}
    \item $\mathcal{O}_i^{\text{SM}}$ is the Standard Model contribution to bin $i$,
    \item $\mathcal{O}_i^{\text{INT}, \bar{t}_R t_R h Z}$ and $\mathcal{O}_i^{\text{INT}, \bar{t}_L t_L h Z}$ are the interference terms between the SM and the EFT operators contributing to $g_{hZt_R}$ and $g_{hZt_L}$ respectively,
    \item $\mathcal{O}_i^{\text{SQ}, \bar{t}_R t_R h Z}$ and $\mathcal{O}_i^{\text{SQ}, \bar{t}_L t_L h Z}$ are the pure squared EFT contributions,
    \item $\mathcal{O}_i^{\text{MIX}}$ denotes the cross term arising from the interference between the $\bar{t}_R t_R h Z$ and $\bar{t}_L t_L h Z$ EFT amplitudes.
\end{itemize}

The bin-wise uncertainty $\sigma_i$ includes both statistical and systematic contributions, and the expected data is assumed to be given by the SM prediction, i.e., $\mathcal{O}_i^{\text{exp.,SM}} = \mathcal{O}_i^{\text{SM}}$. To obtain the bin-wise theoretical predictions for each ${\cal O}_i^X$, we generate multiple samples with different sets of values for $ g_{hZt_R}$ and $g_{hZt_L}$.  This allows us to obtain 
$\chi^2$ as a function of  $ g_{hZt_R}$ and $g_{hZt_L}$.

\begin{figure}[htbp]
  \centering
  
  \includegraphics[width=0.4\textwidth] {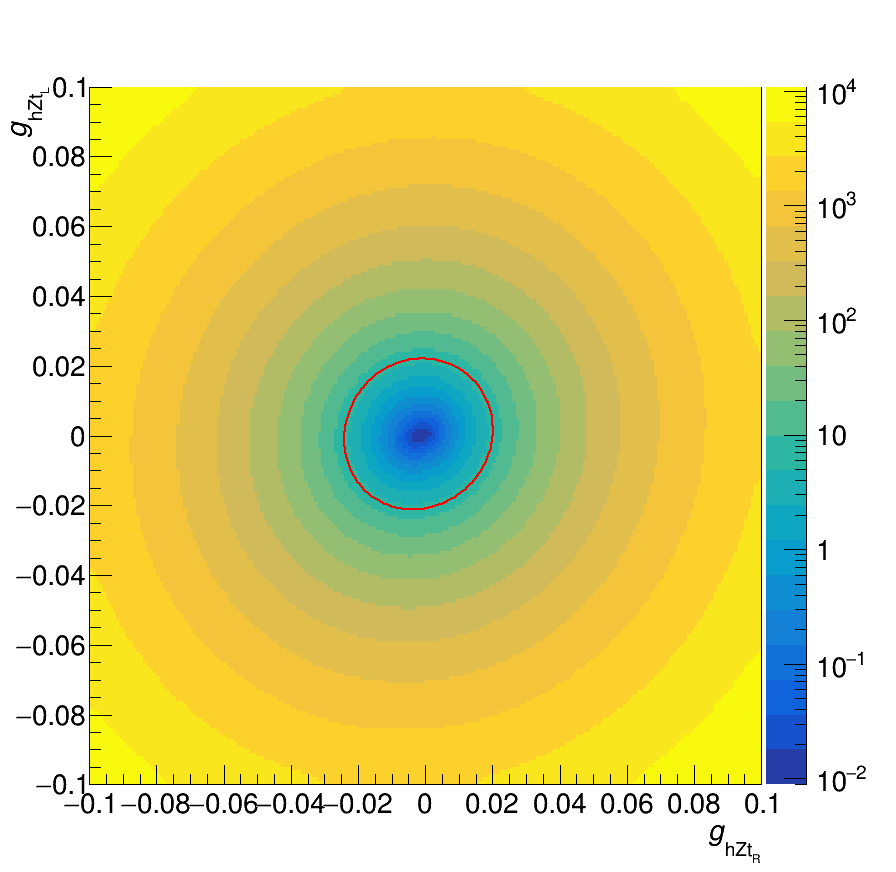}
  \includegraphics[width=0.4\textwidth]{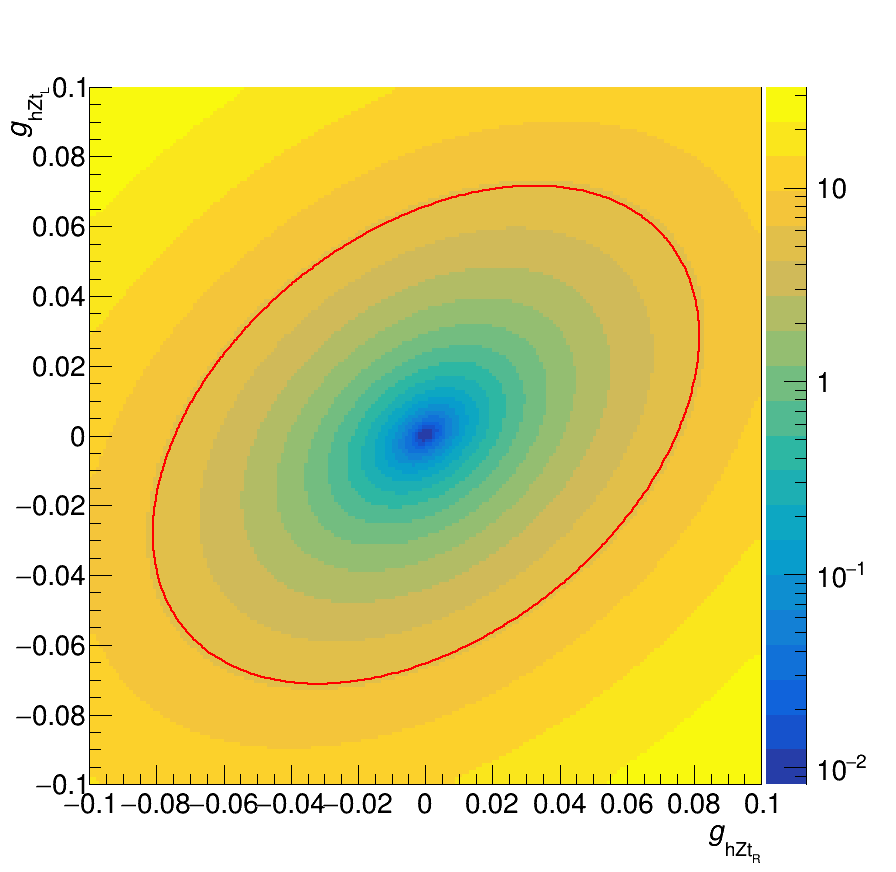}
  \caption{The left panel shows the contours obtained when both the squared and interference contributions are included in the $\chi^{2}$ evaluation, while the right panel displays the result when only the interference terms are retained. The red circle denotes the 95\% C.L. allowed region in the $(g_{hZt_R},, g_{hZt_L})$ plane for an integrated luminosity of $30~\text{ab}^{-1}$. A 5\% systematic uncertainty on the background estimation is included, in addition to the statistical uncertainties beyond 1~TeV. The contour is evaluated using 11 bins: 800 GeV $< m_{hZ} <$  1.5 TeV as shown in Fig.~\ref{fig::reco_misc} (left). The colour axis indicates the $\chi^{2}$ values. In the SMEFT interpretation, these bounds can be translated into constraints on the $Z$-coupling deviations by a simple rescaling of the axes, $\delta g^Z_{t_{L,R}} = g_{hZt_{L,R}}/2$ (see Eq.~\eqref{expression2}).}
  \label{fig::limit2d}
\end{figure}

We then use this $\chi^2$ function to extract the 95\% confidence level bounds in the two-dimensional plane of $g_{hZt_R}$ versus $g_{hZt_L}$, as shown in Fig.~\ref{fig::limit2d}. The 95\% confidence level contour is derived assuming an integrated luminosity of 30~ab$^{-1}$, together with a 5\% systematic uncertainty on the background for $m_{hZ} > 1$~TeV. To evaluate this contour, we restrict to events with $m_{hZ} < 1.5$~TeV, explicitly imposing this cut to ensure EFT validity in accordance with \eq{cutoff}. The colour axis in Fig.~\ref{fig::limit2d} denotes the corresponding $\chi^2$ values.

Within the SMEFT framework, the bounds in Fig.~\ref{fig::limit2d} can be translated to bounds on the $Z$-coupling deviations, $\delta g^Z_{t_{L,R}} =g_{hZt_{L,R}}/2$ (see \eq{expression2}), that can be obtained by a simple rescaling of the axes. This results in percent level bounds competitive with those expected from the 365 GeV run of the FCC-$ee$~\cite{FCC}.\footnote{Note that the $pp \to \bar{t}t Zh$ process is also directly affected by the $Z$-coupling deviations, $\delta g^Z_{t_{L,R}}$. The process, however, is not sensitive to percent level deviations in the $Z\bar{t} t$ coupling. This is because unlike the case of the $hZ\bar{t}t$ contact terms, the corrections to the amplitude due to the $Z$-coupling deviations do not grow quadratically in center of mass energy with respect to the SM (see for eg. Ref.~\cite{Banerjee:2018bio}).}

\section{Summary and Outlook}
\label{sec::Summary}
In this work, we studied the  $pp \to \bar{t}tZh$ process with $Z \to \ell^+\ell^-$ and $h \to b\bar{b}$, at  the Future Circular Collider in its proton–proton configuration (FCC-$hh$). This process is uniquely sensitive to anomalous $\bar{t}tZh$ contact interactions that arise in effective theories like the HEFT and the SMEFT. Given that these couplings are currently unconstrained by existing experiments and will remain poorly probed even at future $e^+e^-$ colliders, the FCC-$hh$ offers a unique opportunity to test this sector of top-quark dynamics. 

To carry out the collider analysis, we developed a dedicated collider analysis for the $4b + 3\ell + \geq 2j + \slashed{E}_T$ final state. Employing realistic detector modelling, including $p_T$ and $\eta$-dependent $b$-tagging and mistag efficiencies, we reconstructed all key intermediate states and analysed the invariant mass and $p_T$ spectra of the $Z$-boson, Higgs boson, and top quark candidates. The interference term of the dimension-6 operators grow quadratically with energy, enhancing the sensitivity of high-mass $Zh$ tails to new physics.

We then performed a binned $\chi^2$ fit over the $m_{Zh}$ distribution to derive the projections in the $g_{hZt_R}$–$g_{hZt_L}$ plane for an integrated luminosity of 30 ab$^{-1}$ at the FCC-$hh$ (see Fig.~\ref{fig::limit2d}). Our results show that values of $g_{hZt_{R/L}}$ down to $\sim{\cal O}(10^{-2})$ can be probed, highlighting the powerful reach of the FCC-$hh$ in exploring top-Higgs electroweak couplings. In the SMEFT, these bounds can be directly translated to percent level bounds on $Z$-coupling deviations that are competitive with FCC-$ee$ sensitivity to probe these deviations~\cite{FCC}.

The analysis presented here represents a first attempt to assess the potential of using the $\bar{t}tZh$ final state to probe further directions in the SMEFT/HEFT space of operators. From the study of the signal and its backgrounds, it is clear that this is extremely challenging even at the FCC-hh. Nevertheless, taking our results as first evidence that interesting constraints are within reach, we believe that further improvements can be made (e.g., relying on machine learning to improve the signal/background separation and assuming better detector performance). Furthermore, future work could extend this analysis by incorporating a global SMEFT/HEFT fit that includes charged-current and other neutral-current processes to fully disentangle correlated directions in the EFT parameter space. This study underscores the role of the FCC-hh in pushing the frontier of top-quark precision measurements and in probing new physics encoded in the SMEFT framework.

\acknowledgments

SJ, EV, and SB acknowledge the \textit{Les Houches workshop series ``Physics at TeV colliders" 2019} where the work was conceived and in the following year, its preliminary version was published as part of the proceedings~\cite{Brooijmans:2020yij}. SB acknowledges the hospitality of the CERN Theory division where this work was completed. SB would also like to thank Durham University IPPP's Batch computing where a major part of the event generation was performed. SJ, and SB acknowledge Elena Gianolio for helping us setup the CERN EOS computing. Finally, SB thanks Daniel Reichelt for some helpful discussions on \texttt{Sherpa}. \\

\flushleft

\textbf{Funding} SB acknowledges support from the Anusandhan National Research Foundation (ANRF), Government of India, under Project ID ANRF/ECRG/2024/004088/PMS, and from the Department of Atomic Energy (DAE), Government of India. RSG acknowledges support from DAE, Government of India, under Project ID RTI 4002. SJ acknowledges support from the DAE, Government of India. \\

\textbf{Data Availability Statement} This manuscript has no associated data. [Author’s comment: The study generated large theoretical datasets that could not be stored in a public repository for practical reasons. However, all datasets, including simulation inputs and run configurations,
are available from the authors upon reasonable request.]

\textbf{Code Availability Statement} This manuscript has no associated code/software. [Author’s comment: The study generated analysis codes that are not yet deposited in a public repository. The authors may make it publicly available in the near future, and in the meantime, it can be shared upon reasonable request.]

\appendix
\section{Dominant EFT contributions to the  $pp \to \bar{t}tZh$ process for different $Z$ polarisations.}
\label{amplitude}
In this Appendix we discuss the amplitude for the EFT corrections due  to the $hZtt$ contact term,
\bea
V_\mu^{ij}= \frac{1}{2}\left((g^{h}_{Zt_R}+g^{h}_{Zt_L})\gamma^{ij}_\mu+(g^{h}_{Zt_R}-g^{h}_{Zt_L})(\gamma_\mu.\gamma_5)^{ij}\right)
\eea
where the indices $i,j$ denote the rows and columns of the gamma matrices.  The relevant diagrams can be obtained from the Feynman diagrams in the second row of Fig.~\ref{fig:tthz_diagrams} by replacing one of the red blobs by the two diagrams shown in the legend. The first diagram in the legend shows the SM contribution whereas the second one gives the correction due to the EFT vertex in Eq.~\ref{anam}. Replacing one of the red blobs in each of the diagrams in the second row gives us 9 contributions.  The amplitude for the $I$-th contribution can be written as, 
\bea
{\cal M}^I={\cal M}^{I}_{ij} \times \left((v \gamma_{\mu}^{ij} +a(\gamma_5.\gamma_{\mu})^{ij} ) \frac{g^{\mu \nu}}{M_{Zh}^2-m_Z^2} +V^{\nu, ij}\right) \epsilon^*_{Z,\nu}
\eea
where ${\cal M}^{I}_{ij}$ is the contribution from the rest of the diagram excluding the blob; $v$ and $a$ are the SM vector  and axial couplings of the $Z$-boson; and $\epsilon^*_{Z,\nu}$ is the polarisation vector of the $Z$ boson. Due to the absence of the $Z$ propagator the EFT contribution, the contact terms grow quadratically in $M_{Zh}$ with respect to the SM contribution. Once the amplitude is squared the interference term will thus grow quadratically in $M_{Zh}$ with respect to  the SM squared contribution. This is true for each of the $Z$ polarisations. The energy-growing behaviour can be understood from the structure of the polarisation vectors. In the high-energy limit, the longitudinal polarisation vector scales as $\epsilon_L^\mu \sim p^\mu/m_V$, leading to an explicit energy enhancement in the amplitude, while the transverse polarisations satisfy $\epsilon_T^\mu=\mathcal{O}(1)$. Consequently, deviations from the SM induce a more pronounced energy growth in the longitudinal mode than in the transverse one.

\bibliographystyle{JHEP}
\bibliography{refs}

@article{Denner:2015fca,
    author = "Denner, Ansgar and Dittmaier, Stefan and Hecht, Markus and Pasold, Christian",
    title = "{NLO QCD and electroweak corrections to $Z+\gamma$ production with leptonic Z-boson decays}",
    eprint = "1510.08742",
    archivePrefix = "arXiv",
    primaryClass = "hep-ph",
    reportNumber = "FR-PHENO-2015-012",
    doi = "10.1007/JHEP02(2016)057",
    journal = "JHEP",
    volume = "02",
    pages = "057",
    year = "2016"
}

@article{Kallweit:2015dum,
    author = {Kallweit, Stefan and Lindert, Jonas M. and Maierhofer, Philipp and Pozzorini, Stefano and Sch{\"o}nherr, Marek},
    title = "{NLO QCD+EW predictions for V + jets including off-shell vector-boson decays and multijet merging}",
    eprint = "1511.08692",
    archivePrefix = "arXiv",
    primaryClass = "hep-ph",
    reportNumber = "DCPT-15-140, FR-PHENO-2015-014, IPPP-15-70, MCNET-15-23, ZU-TH-41-15, MITP-15-108",
    doi = "10.1007/JHEP04(2016)021",
    journal = "JHEP",
    volume = "04",
    pages = "021",
    year = "2016"
}

@article{Denner:2000jv,
    author = "Denner, Ansgar and Pozzorini, Stefano",
    title = "{One loop leading logarithms in electroweak radiative corrections. 1. Results}",
    eprint = "hep-ph/0010201",
    archivePrefix = "arXiv",
    reportNumber = "PSI-PR-00-15, ZU-TH-17-00",
    doi = "10.1007/s100520100551",
    journal = "Eur. Phys. J. C",
    volume = "18",
    pages = "461--480",
    year = "2001"
}

@article{Denner:2001gw,
    author = "Denner, Ansgar and Pozzorini, Stefano",
    title = "{One loop leading logarithms in electroweak radiative corrections. 2. Factorization of collinear singularities}",
    eprint = "hep-ph/0104127",
    archivePrefix = "arXiv",
    reportNumber = "PSI-PR-01-06, ZU-TH-13-01",
    doi = "10.1007/s100520100721",
    journal = "Eur. Phys. J. C",
    volume = "21",
    pages = "63--79",
    year = "2001"
}

@article{Ciafaloni:2000df,
    author = "Ciafaloni, Marcello and Ciafaloni, Paolo and Comelli, Denis",
    title = "{Bloch-Nordsieck violating electroweak corrections to inclusive TeV scale hard processes}",
    eprint = "hep-ph/0001142",
    archivePrefix = "arXiv",
    doi = "10.1103/PhysRevLett.84.4810",
    journal = "Phys. Rev. Lett.",
    volume = "84",
    pages = "4810--4813",
    year = "2000"
}

@article{Melles:2001dh,
    author = "Melles, Michael",
    title = "{Resummation of angular dependent corrections in spontaneously broken gauge theories}",
    eprint = "hep-ph/0108221",
    archivePrefix = "arXiv",
    reportNumber = "PSI-PR-01-11",
    doi = "10.1007/s100520200942",
    journal = "Eur. Phys. J. C",
    volume = "24",
    pages = "193--204",
    year = "2002"
}

@article{Banerjee:2024eyo,
    author = "Banerjee, Shankha and Reichelt, Daniel and Spannowsky, Michael",
    title = "{Electroweak corrections and EFT operators in W+W- production at the LHC}",
    eprint = "2406.15640",
    archivePrefix = "arXiv",
    primaryClass = "hep-ph",
    reportNumber = "IPPP/24/34, MCNET-24-11, IPPP/24/34 MCNET-24-11",
    doi = "10.1103/PhysRevD.110.115012",
    journal = "Phys. Rev. D",
    volume = "110",
    number = "11",
    pages = "115012",
    year = "2024"
}

@article{Aguilar-Saavedra:2014iga,
    author = "Aguilar-Saavedra, Juan A. and Fuks, Benjamin and Mangano, Michelangelo L.",
    title = "{Pinning down top dipole moments with ultra-boosted tops}",
    eprint = "1412.6654",
    archivePrefix = "arXiv",
    primaryClass = "hep-ph",
    reportNumber = "CERN-PH-TH-2014-259",
    doi = "10.1103/PhysRevD.91.094021",
    journal = "Phys. Rev. D",
    volume = "91",
    pages = "094021",
    year = "2015"
}

@article{Gupta:2014rxa,
    author = "Gupta, Rick S. and Pomarol, Alex and Riva, Francesco",
    title = "{BSM Primary Effects}",
    eprint = "1405.0181",
    archivePrefix = "arXiv",
    primaryClass = "hep-ph",
    doi = "10.1103/PhysRevD.91.035001",
    journal = "Phys. Rev. D",
    volume = "91",
    number = "3",
    pages = "035001",
    year = "2015"
}

@article{silh,
    author = "Giudice, G. F. and Grojean, C. and Pomarol, A. and Rattazzi, R.",
    title = "{The Strongly-Interacting Light Higgs}",
    eprint = "hep-ph/0703164",
    archivePrefix = "arXiv",
    reportNumber = "CERN-PH-TH-2007-47",
    doi = "10.1088/1126-6708/2007/06/045",
    journal = "JHEP",
    volume = "06",
    pages = "045",
    year = "2007"
}

@article{FCC,
    author = "Benedikt, M. and others",
    collaboration = "FCC",
    title = "{Future Circular Collider Feasibility Study Report: Volume 1, Physics, Experiments, Detectors}",
    eprint = "2505.00272",
    archivePrefix = "arXiv",
    primaryClass = "hep-ex",
    reportNumber = "CERN-FCC-PHYS-2025-0002",
    doi = "10.17181/CERN.9DKX.TDH9",
    month = "4",
    year = "2025"
}

@article{grojeanww,
    author = "Grojean, Christophe and Montull, Marc and Riembau, Marc",
    title = "{Diboson at the LHC vs LEP}",
    eprint = "1810.05149",
    archivePrefix = "arXiv",
    primaryClass = "hep-ph",
    reportNumber = "DESY 17-231, DESY-17-231",
    doi = "10.1007/JHEP03(2019)020",
    journal = "JHEP",
    volume = "03",
    pages = "020",
    year = "2019"
}

@article{pomarolw,
    author = "Franceschini, Roberto and Panico, Giuliano and Pomarol, Alex and Riva, Francesco and Wulzer, Andrea",
    title = "{Electroweak Precision Tests in High-Energy Diboson Processes}",
    eprint = "1712.01310",
    archivePrefix = "arXiv",
    primaryClass = "hep-ph",
    reportNumber = "CERN-TH-2017-252, RM3-TH-17-1",
    doi = "10.1007/JHEP02(2018)111",
    journal = "JHEP",
    volume = "02",
    pages = "111",
    year = "2018"
}

@article{FCC:2025uan,
    author = "Benedikt, M. and others",
    collaboration = "FCC",
    title = "{Future Circular Collider Feasibility Study Report: Volume 2, Accelerators, Technical Infrastructure and Safety}",
    eprint = "2505.00274",
    archivePrefix = "arXiv",
    primaryClass = "physics.acc-ph",
    reportNumber = "CERN-FCC-ACC-2025-0004",
    doi = "10.1140/epjs/s11734-025-01967-4",
    journal = "Eur. Phys. J. ST",
    volume = "234",
    number = "19",
    pages = "5713--6197",
    year = "2025"
}

@inproceedings{Brooijmans:2020yij,
    author = "Brooijmans, G. and others",
    title = "{Les Houches 2019 Physics at TeV Colliders: New Physics Working Group Report}",
    booktitle = "{11th Les Houches Workshop on Physics at TeV Colliders}: {PhysTeV Les Houches}",
    eprint = "2002.12220",
    archivePrefix = "arXiv",
    primaryClass = "hep-ph",
    month = "2",
    year = "2020"
}

@article{Amar:2014fpa,
    author = "Amar, Gilad and Banerjee, Shankha and von Buddenbrock, Stefan and Cornell, Alan S. and Mandal, Tanumoy and Mellado, Bruce and Mukhopadhyaya, Biswarup",
title = {Exploration of the tensor structure of the Higgs boson coupling to weak bosons in $e^+ e^-$ collisions},
    eprint = "1405.3957",
    archivePrefix = "arXiv",
    primaryClass = "hep-ph",
    reportNumber = "HRI-RECAPP-2014-011, WITS-CTP-135",
    doi = "10.1007/JHEP02(2015)128",
    journal = "JHEP",
    volume = "02",
    pages = "128",
    year = "2015"
}

@article{Banerjee:2015bla,
    author = "Banerjee, Shankha and Mandal, Tanumoy and Mellado, Bruce and Mukhopadhyaya, Biswarup",
    title = "{Cornering dimension-6 $HVV$ interactions at high luminosity LHC: the role of event ratios}",
    eprint = "1505.00226",
    archivePrefix = "arXiv",
    primaryClass = "hep-ph",
    reportNumber = "HRI-RECAPP-2015-008, WITS-MITP-006",
    doi = "10.1007/JHEP09(2015)057",
    journal = "JHEP",
    volume = "09",
    pages = "057",
    year = "2015"
}

@article{Falkowski:2014tna,
    author = "Falkowski, Adam and Riva, Francesco",
    title = "{Model-independent precision constraints on dimension-6 operators}",
    eprint = "1411.0669",
    archivePrefix = "arXiv",
    primaryClass = "hep-ph",
    reportNumber = "LPT-ORSAY-14-77",
    doi = "10.1007/JHEP02(2015)039",
    journal = "JHEP",
    volume = "02",
    pages = "039",
    year = "2015"
}

@article{Banerjee:2019twi,
    author = "Banerjee, Shankha and Gupta, Rick S. and Reiness, Joey Y. and Seth, Satyajit and Spannowsky, Michael",
    title = "{Towards the ultimate differential SMEFT analysis}",
    eprint = "1912.07628",
    archivePrefix = "arXiv",
    primaryClass = "hep-ph",
    reportNumber = "IPPP/19/93",
    doi = "10.1007/JHEP09(2020)170",
    journal = "JHEP",
    volume = "09",
    pages = "170",
    year = "2020"
}

@article{Banerjee:2019pks,
    author = "Banerjee, Shankha and Gupta, Rick S. and Reiness, Joey Y. and Spannowsky, Michael",
    title = "{Resolving the tensor structure of the Higgs coupling to $Z$-bosons via Higgs-strahlung}",
    eprint = "1905.02728",
    archivePrefix = "arXiv",
    primaryClass = "hep-ph",
    reportNumber = "IPPP/19/35",
    doi = "10.1103/PhysRevD.100.115004",
    journal = "Phys. Rev. D",
    volume = "100",
    number = "11",
    pages = "115004",
    year = "2019"
}

@article{Banerjee:2018bio,
    author = "Banerjee, Shankha and Englert, Christoph and Gupta, Rick S. and Spannowsky, Michael",
    title = "{Probing Electroweak Precision Physics via boosted Higgs-strahlung at the LHC}",
    eprint = "1807.01796",
    archivePrefix = "arXiv",
    primaryClass = "hep-ph",
    reportNumber = "IPPP/18/53, IPPP-18-53",
    doi = "10.1103/PhysRevD.98.095012",
    journal = "Phys. Rev. D",
    volume = "98",
    number = "9",
    pages = "095012",
    year = "2018"
}

@article{Banerjee:2021huv,
    author = "Banerjee, Shankha and Gupta, Rick S. and Ochoa-Valeriano, Oscar and Spannowsky, Michael",
    title = "{High energy lepton colliders as the ultimate Higgs microscopes}",
    eprint = "2109.14634",
    archivePrefix = "arXiv",
    primaryClass = "hep-ph",
    reportNumber = "CERN-TH-2021-142, IPPP/21/35",
    doi = "10.1007/JHEP02(2022)176",
    journal = "JHEP",
    volume = "02",
    pages = "176",
    year = "2022"
}

@article{Bishara:2022vsc,
    author = "Bishara, Fady and Englert, Philipp and Grojean, Christophe and Panico, Giuliano and Rossia, Alejo N.",
    title = "{Revisiting Vh(\textrightarrow{}$ b\overline{b} $) at the LHC and FCC-hh}",
    eprint = "2208.11134",
    archivePrefix = "arXiv",
    primaryClass = "hep-ph",
    reportNumber = "DESY 22-136, HU-EP-22/27",
    doi = "10.1007/JHEP06(2023)077",
    journal = "JHEP",
    volume = "06",
    pages = "077",
    year = "2023"
}

@article{Araz:2020zyh,
    author = "Araz, Jack Y. and Banerjee, Shankha and Gupta, Rick S. and Spannowsky, Michael",
    title = "{Precision SMEFT bounds from the VBF Higgs at high transverse momentum}",
    eprint = "2011.03555",
    archivePrefix = "arXiv",
    primaryClass = "hep-ph",
    reportNumber = "IPPP/20/52, CERN-TH-2020-186",
    doi = "10.1007/JHEP04(2021)125",
    journal = "JHEP",
    volume = "04",
    pages = "125",
    year = "2021"
}

@article{Dror:2015nkp,
    author = "Dror, Jeff Asaf and Farina, Marco and Salvioni, Ennio and Serra, Javi",
    title = "{Strong tW Scattering at the LHC}",
    eprint = "1511.03674",
    archivePrefix = "arXiv",
    primaryClass = "hep-ph",
    reportNumber = "CERN-PH-TH-2015-265",
    doi = "10.1007/JHEP01(2016)071",
    journal = "JHEP",
    volume = "01",
    pages = "071",
    year = "2016"
}

@article{Alloul:2013bka,
    author = "Alloul, Adam and Christensen, Neil D. and Degrande, C\'eline and Duhr, Claude and Fuks, Benjamin",
    title = "{FeynRules  2.0 - A complete toolbox for tree-level phenomenology}",
    eprint = "1310.1921",
    archivePrefix = "arXiv",
    primaryClass = "hep-ph",
    reportNumber = "CERN-PH-TH-2013-239, MCNET-13-14, IPPP-13-71, DCPT-13-142, PITT-PACC-1308",
    doi = "10.1016/j.cpc.2014.04.012",
    journal = "Comput. Phys. Commun.",
    volume = "185",
    pages = "2250--2300",
    year = "2014"
}

@article{Degrande:2011ua,
    author = "Degrande, Celine and Duhr, Claude and Fuks, Benjamin and Grellscheid, David and Mattelaer, Olivier and Reiter, Thomas",
    title = "{UFO - The Universal FeynRules Output}",
    eprint = "1108.2040",
    archivePrefix = "arXiv",
    primaryClass = "hep-ph",
    reportNumber = "CP3-11-25, IPHC-PHENO-11-04, IPPP-11-39, DCPT-11-78, MPP-2011-68",
    doi = "10.1016/j.cpc.2012.01.022",
    journal = "Comput. Phys. Commun.",
    volume = "183",
    pages = "1201--1214",
    year = "2012"
}

@article{Sherpa:2019gpd,
    author = "Bothmann, Enrico and others",
    collaboration = "Sherpa",
    title = "{Event Generation with Sherpa 2.2}",
    eprint = "1905.09127",
    archivePrefix = "arXiv",
    primaryClass = "hep-ph",
    reportNumber = "FERMILAB-PUB-19-218-T, SLAC-PUB-17433, IPPP/19/42, MCNET-19-11",
    doi = "10.21468/SciPostPhys.7.3.034",
    journal = "SciPost Phys.",
    volume = "7",
    number = "3",
    pages = "034",
    year = "2019"
}

@article{Gleisberg:2008fv,
    author = "Gleisberg, Tanju and Hoeche, Stefan",
    title = "{Comix, a new matrix element generator}",
    eprint = "0808.3674",
    archivePrefix = "arXiv",
    primaryClass = "hep-ph",
    reportNumber = "SLAC-PUB-13232, IPPP-08-31, DCPT-08-62, MCNET-08-08",
    doi = "10.1088/1126-6708/2008/12/039",
    journal = "JHEP",
    volume = "12",
    pages = "039",
    year = "2008"
}

@article{Buckley:2014ana,
    author = {Buckley, Andy and Ferrando, James and Lloyd, Stephen and Nordstr\"om, Karl and Page, Ben and R\"ufenacht, Martin and Sch\"onherr, Marek and Watt, Graeme},
    title = "{LHAPDF6: parton density access in the LHC precision era}",
    eprint = "1412.7420",
    archivePrefix = "arXiv",
    primaryClass = "hep-ph",
    reportNumber = "GLAS-PPE-2014-05, MCNET-14-29, IPPP-14-111, DCPT-14-222",
    doi = "10.1140/epjc/s10052-015-3318-8",
    journal = "Eur. Phys. J. C",
    volume = "75",
    pages = "132",
    year = "2015"
}

@article{Catani:2001cc,
  author    = "Catani, Stefano and Krauss, Frank and Kuhn, R. and Webber, Bryan R.",
  title     = "{QCD matrix elements + parton showers}",
  journal   = "JHEP",
  volume    = "11",
  pages     = "063",
  year      = "2001",
  eprint    = "hep-ph/0109231",
  archivePrefix = "arXiv",
  reportNumber = "Cavendish-HEP-01/12, IPPP-01-56, DCPT-01-112",
  doi       = "10.1088/1126-6708/2001/11/063"
}

@article{Cacciari:2008gp,
    author = "Cacciari, Matteo and Salam, Gavin P. and Soyez, Gregory",
    title = "{The anti-$k_t$ jet clustering algorithm}",
    eprint = "0802.1189",
    archivePrefix = "arXiv",
    primaryClass = "hep-ph",
    reportNumber = "LPTHE-07-03",
    doi = "10.1088/1126-6708/2008/04/063",
    journal = "JHEP",
    volume = "04",
    pages = "063",
    year = "2008"
}

@article{Cacciari:2011ma,
    author = "Cacciari, Matteo and Salam, Gavin P. and Soyez, Gregory",
    title = "{FastJet User Manual}",
    eprint = "1111.6097",
    archivePrefix = "arXiv",
    primaryClass = "hep-ph",
    reportNumber = "CERN-PH-TH-2011-297",
    doi = "10.1140/epjc/s10052-012-1896-2",
    journal = "Eur. Phys. J. C",
    volume = "72",
    pages = "1896",
    year = "2012"
}

@article{Buckley:2019xhk,
    author = {Buckley, Andy and Ilten, Philip and Konstantinov, Dmitri and L\"onnblad, Leif and Monk, James and Pokorski, Witold and Przedzinski, Tomasz and Verbytskyi, Andrii},
    title = "{The HepMC3 event record library for Monte Carlo event generators}",
    eprint = "1912.08005",
    archivePrefix = "arXiv",
    primaryClass = "hep-ph",
    reportNumber = "MPP-2019-258, MCNET-19-27, LU-TP 19-58",
    doi = "10.1016/j.cpc.2020.107310",
    journal = "Comput. Phys. Commun.",
    volume = "260",
    pages = "107310",
    year = "2021"
}

@techreport{Selvaggi:2717698,
      author        = "Selvaggi, Michele",
      title         = "{A Delphes parameterisation of the FCC-hh detector}",
      institution   = "CERN",
      reportNumber  = "CERN-FCC-PHYS-2020-0003",
      address       = "Geneva",
      year          = "2020",
      url           = "https://cds.cern.ch/record/2717698",
}

@article{Hernandez-Juarez:2018uow,
    author = "Hern\'andez-Ju\'arez, A. I. and Moyotl, A. and Tavares-Velasco, G.",
    title = "{Chromomagnetic and chromoelectric dipole moments of the top quark in the fourth-generation THDM}",
    eprint = "1805.00615",
    archivePrefix = "arXiv",
    primaryClass = "hep-ph",
    doi = "10.1103/PhysRevD.98.035040",
    journal = "Phys. Rev. D",
    volume = "98",
    number = "3",
    pages = "035040",
    year = "2018"
}

@article{Gaitan:2015aia,
    author = "Gaitan, R. and Garces, E. A. and de Oca, J. H. Montes and Martinez, R.",
    title = "{Top quark Chromoelectric and Chromomagnetic Dipole Moments in a Two Higgs Doublet Model with CP violation}",
    eprint = "1505.04168",
    archivePrefix = "arXiv",
    primaryClass = "hep-ph",
    doi = "10.1103/PhysRevD.92.094025",
    journal = "Phys. Rev. D",
    volume = "92",
    number = "9",
    pages = "094025",
    year = "2015"
}

@article{Kidonakis:2023htm,
    author = "Kidonakis, Nikolaos and Tonero, Alberto",
    title = "{SMEFT chromomagnetic dipole operator contributions to $t{{\bar{t}}}$ production at approximate NNLO in QCD}",
    eprint = "2309.16758",
    archivePrefix = "arXiv",
    primaryClass = "hep-ph",
    doi = "10.1140/epjc/s10052-024-12938-9",
    journal = "Eur. Phys. J. C",
    volume = "84",
    number = "6",
    pages = "591",
    year = "2024"
}

@article{Khatibi:2020mvt,
    author = "Khatibi, Sara and Khanpour, Hamzeh",
    title = "{Probing four-fermion operators in the triple top production at future hadron colliders}",
    eprint = "2011.15060",
    archivePrefix = "arXiv",
    primaryClass = "hep-ph",
    doi = "10.1016/j.nuclphysb.2021.115432",
    journal = "Nucl. Phys. B",
    volume = "967",
    pages = "115432",
    year = "2021"
}

@article{DHondt:2018cww,
    author = "D'Hondt, Jorgen and Mariotti, Alberto and Mimasu, Ken and Moortgat, Seth and Zhang, Cen",
    title = "{Learning to pinpoint effective operators at the LHC: a study of the $ \mathrm{t}\overline{\mathrm{t}}\mathrm{b}\overline{\mathrm{b}} $ signature}",
    eprint = "1807.02130",
    archivePrefix = "arXiv",
    primaryClass = "hep-ph",
    reportNumber = "CP3-18-42",
    doi = "10.1007/JHEP11(2018)131",
    journal = "JHEP",
    volume = "11",
    pages = "131",
    year = "2018"
}

@article{Aguilar-Saavedra:2010uur,
    author = "Aguilar-Saavedra, J. A.",
    title = "{Effective four-fermion operators in top physics: A Roadmap}",
    eprint = "1008.3562",
    archivePrefix = "arXiv",
    primaryClass = "hep-ph",
    doi = "10.1016/j.nuclphysb.2011.06.003",
    journal = "Nucl. Phys. B",
    volume = "843",
    pages = "638--672",
    year = "2011",
    note = "[Erratum: Nucl.Phys.B 851, 443--444 (2011)]"
}

@article{Brivio:2019ius,
    author = "Brivio, Ilaria and Bruggisser, Sebastian and Maltoni, Fabio and Moutafis, Rhea and Plehn, Tilman and Vryonidou, Eleni and Westhoff, Susanne and Zhang, C.",
    title = "{O new physics, where art thou? A global search in the top sector}",
    eprint = "1910.03606",
    archivePrefix = "arXiv",
    primaryClass = "hep-ph",
    reportNumber = "P3H-19-036, CERN-TH-2019-193",
    doi = "10.1007/JHEP02(2020)131",
    journal = "JHEP",
    volume = "02",
    pages = "131",
    year = "2020"
}

@article{DiNoi:2023ygk,
    author = {Di Noi, Stefano and Gr\"ober, Ramona and Heinrich, Gudrun and Lang, Jannis and Vitti, Marco},
    title = "{\ensuremath{\gamma}5 schemes and the interplay of SMEFT operators in the Higgs-gluon coupling}",
    eprint = "2310.18221",
    archivePrefix = "arXiv",
    primaryClass = "hep-ph",
    reportNumber = "P3H-23-080, KA-TP-22-2023, TTP23-054",
    doi = "10.1103/PhysRevD.109.095024",
    journal = "Phys. Rev. D",
    volume = "109",
    number = "9",
    pages = "095024",
    year = "2024"
}

@article{Bhattacharya:2022kje,
    author = "Bhattacharya, Subhaditya and Biswas, Sanjoy and Pal, Kuntal and Wudka, Jose",
    title = "{Associated production of Higgs and single top at the LHC in presence of the SMEFT operators}",
    eprint = "2211.05450",
    archivePrefix = "arXiv",
    primaryClass = "hep-ph",
    doi = "10.1007/JHEP08(2023)015",
    journal = "JHEP",
    volume = "08",
    pages = "015",
    year = "2023"
}

@article{DiNoi:2023onw,
    author = {Di Noi, Stefano and Gr\"ober, Ramona},
    title = "{Renormalisation group running effects in $pp\rightarrow t{\bar{t}}h$ in the Standard Model Effective Field Theory}",
    eprint = "2312.11327",
    archivePrefix = "arXiv",
    primaryClass = "hep-ph",
    doi = "10.1140/epjc/s10052-024-12661-5",
    journal = "Eur. Phys. J. C",
    volume = "84",
    number = "4",
    pages = "403",
    year = "2024"
}

@article{Barger:2023wbg,
    author = "Barger, Vernon and Hagiwara, Kaoru and Zheng, Ya-Juan",
    title = "{CP-violating top-Higgs coupling in SMEFT}",
    eprint = "2310.10852",
    archivePrefix = "arXiv",
    primaryClass = "hep-ph",
    reportNumber = "KEK-TH-2559",
    doi = "10.1016/j.physletb.2024.138547",
    journal = "Phys. Lett. B",
    volume = "850",
    pages = "138547",
    year = "2024"
}

@article{Bhardwaj:2023ufl,
    author = "Bhardwaj, Akanksha and Englert, Christoph and Gon\c{c}alves, Dorival and Navarro, Alberto",
    title = "{Nonlinear CP violation in the top-Higgs sector}",
    eprint = "2308.11722",
    archivePrefix = "arXiv",
    primaryClass = "hep-ph",
    doi = "10.1103/PhysRevD.108.115006",
    journal = "Phys. Rev. D",
    volume = "108",
    number = "11",
    pages = "115006",
    year = "2023"
}

@article{Banerjee:2019jys,
    author = "Banerjee, Shankha and Krauss, Frank and Spannowsky, Michael",
    title = "{Revisiting the $t\bar{t}hh$ channel at the FCC-hh}",
    eprint = "1904.07886",
    archivePrefix = "arXiv",
    primaryClass = "hep-ph",
    reportNumber = "IPPP/19/29",
    doi = "10.1103/PhysRevD.100.073012",
    journal = "Phys. Rev. D",
    volume = "100",
    pages = "073012",
    year = "2019"
}

@article{Englert:2014uqa,
    author = "Englert, Christoph and Krauss, Frank and Spannowsky, Michael and Thompson, Jennifer",
    title = "{Di-Higgs phenomenology in $t\bar{t}hh$: The forgotten channel}",
    eprint = "1409.8074",
    archivePrefix = "arXiv",
    primaryClass = "hep-ph",
    reportNumber = "IPPP-14-82, DCPT-14-164",
    doi = "10.1016/j.physletb.2015.02.041",
    journal = "Phys. Lett. B",
    volume = "743",
    pages = "93--97",
    year = "2015"
}

@article{Alioli:2017ces,
    author = "Alioli, S. and Cirigliano, V. and Dekens, W. and de Vries, J. and Mereghetti, E.",
    title = "{Right-handed charged currents in the era of the Large Hadron Collider}",
    eprint = "1703.04751",
    archivePrefix = "arXiv",
    primaryClass = "hep-ph",
    reportNumber = "CERN-TH-2017-043, LA-UR-17-21898, NIKHEF-2017-014",
    doi = "10.1007/JHEP05(2017)086",
    journal = "JHEP",
    volume = "05",
    pages = "086",
    year = "2017"
}

@article{Lillie:2007hd,
    author = "Lillie, Ben and Shu, Jing and Tait, Timothy M. P.",
    title = "{Top Compositeness at the Tevatron and LHC}",
    eprint = "0712.3057",
    archivePrefix = "arXiv",
    primaryClass = "hep-ph",
    reportNumber = "ANL-HEP-PR-07-96, EFI-07-31, NUHEP-TH-07-11",
    doi = "10.1088/1126-6708/2008/04/087",
    journal = "JHEP",
    volume = "04",
    pages = "087",
    year = "2008"
}

@article{Pomarol:2008bh,
    author = "Pomarol, Alex and Serra, Javi",
    title = "{Top Quark Compositeness: Feasibility and Implications}",
    eprint = "0806.3247",
    archivePrefix = "arXiv",
    primaryClass = "hep-ph",
    reportNumber = "UAB-FT-647",
    doi = "10.1103/PhysRevD.78.074026",
    journal = "Phys. Rev. D",
    volume = "78",
    pages = "074026",
    year = "2008"
}

@article{Darme:2021gtt,
    author = "Darm\'e, Luc and Fuks, Benjamin and Maltoni, Fabio",
    title = "{Top-philic heavy resonances in four-top final states and their EFT interpretation}",
    eprint = "2104.09512",
    archivePrefix = "arXiv",
    primaryClass = "hep-ph",
    doi = "10.1007/JHEP09(2021)143",
    journal = "JHEP",
    volume = "09",
    pages = "143",
    year = "2021"
}

@article{Malkawi:1994tg,
    author = "Malkawi, Ehab and Yuan, C. P.",
    title = "{A Global analysis of the top quark couplings to gauge bosons}",
    eprint = "hep-ph/9405322",
    archivePrefix = "arXiv",
    reportNumber = "MSUHEP-94-06",
    doi = "10.1103/PhysRevD.50.4462",
    journal = "Phys. Rev. D",
    volume = "50",
    pages = "4462--4477",
    year = "1994"
}

@article{Spira:1997ce,
    author = "Spira, Michael and Wells, James D.",
    title = "{Higgs bosons strongly coupled to the top quark}",
    eprint = "hep-ph/9711410",
    archivePrefix = "arXiv",
    reportNumber = "SLAC-PUB-7703, CERN-TH-97-331",
    doi = "10.1016/S0550-3213(98)00107-2",
    journal = "Nucl. Phys. B",
    volume = "523",
    pages = "3--16",
    year = "1998"
}

@article{CMS:2022quh,
    author = "Tumasyan, Armen and others",
    collaboration = "CMS",
    title = "{Search for CP violating top quark couplings in pp collisions at $ \sqrt{s} $ = 13 TeV}",
    eprint = "2205.07434",
    archivePrefix = "arXiv",
    primaryClass = "hep-ex",
    reportNumber = "CMS-TOP-18-007, CERN-EP-2021-143",
    doi = "10.1007/JHEP07(2023)023",
    journal = "JHEP",
    volume = "07",
    pages = "023",
    year = "2023"
}

@article{Gupta:2009wu,
    author = "Gupta, Sudhir Kumar and Mete, Alaettin Serhan and Valencia, G.",
    title = "{CP violating anomalous top-quark couplings at the LHC}",
    eprint = "0905.1074",
    archivePrefix = "arXiv",
    primaryClass = "hep-ph",
    doi = "10.1103/PhysRevD.80.034013",
    journal = "Phys. Rev. D",
    volume = "80",
    pages = "034013",
    year = "2009"
}

@article{BessidskaiaBylund:2016jvp,
    author = "Bessidskaia Bylund, Olga and Maltoni, Fabio and Tsinikos, Ioannis and Vryonidou, Eleni and Zhang, Cen",
    title = "{Probing top quark neutral couplings in the Standard Model Effective Field Theory at NLO in QCD}",
    eprint = "1601.08193",
    archivePrefix = "arXiv",
    primaryClass = "hep-ph",
    reportNumber = "CP3-16-03, MCNET-16-03",
    doi = "10.1007/JHEP05(2016)052",
    journal = "JHEP",
    volume = "05",
    pages = "052",
    year = "2016"
}

@article{Contino:2006qr,
    author = "Contino, Roberto and Da Rold, Leandro and Pomarol, Alex",
    title = "{Light custodians in natural composite Higgs models}",
    eprint = "hep-ph/0612048",
    archivePrefix = "arXiv",
    reportNumber = "UAB-FT-619, ROMA1-1445-2006",
    doi = "10.1103/PhysRevD.75.055014",
    journal = "Phys. Rev. D",
    volume = "75",
    pages = "055014",
    year = "2007"
}

@article{Matsedonskyi:2012ym,
    author = "Matsedonskyi, Oleksii and Panico, Giuliano and Wulzer, Andrea",
    title = "{Light Top Partners for a Light Composite Higgs}",
    eprint = "1204.6333",
    archivePrefix = "arXiv",
    primaryClass = "hep-ph",
    doi = "10.1007/JHEP01(2013)164",
    journal = "JHEP",
    volume = "01",
    pages = "164",
    year = "2013"
}

@article{Grojean:2004xa,
    author = "Grojean, Christophe and Servant, Geraldine and Wells, James D.",
    title = "{First-order electroweak phase transition in the standard model with a low cutoff}",
    eprint = "hep-ph/0407019",
    archivePrefix = "arXiv",
    reportNumber = "SACLAY-T04-084, MCTP-04-37, ANL-HEP-PR-04-63, EFI-04-23",
    doi = "10.1103/PhysRevD.71.036001",
    journal = "Phys. Rev. D",
    volume = "71",
    pages = "036001",
    year = "2005"
}

@article{Redi:2012ha,
    author = "Redi, Michele and Tesi, Andrea",
    title = "{Implications of a Light Higgs in Composite Models}",
    eprint = "1205.0232",
    archivePrefix = "arXiv",
    primaryClass = "hep-ph",
    reportNumber = "CERN-PH-TH-2012-107",
    doi = "10.1007/JHEP10(2012)166",
    journal = "JHEP",
    volume = "10",
    pages = "166",
    year = "2012"
}

@book{Panico:2015jxa,
    author = "Panico, Giuliano and Wulzer, Andrea",
    title = "{The Composite Nambu-Goldstone Higgs}",
    eprint = "1506.01961",
    archivePrefix = "arXiv",
    primaryClass = "hep-ph",
    reportNumber = "DFPD-2015TH9",
    doi = "10.1007/978-3-319-22617-0",
    publisher = "Springer",
    volume = "913",
    year = "2016"
}

@article{Buckley:2015lku,
    author = "Buckley, Andy and Englert, Christoph and Ferrando, James and Miller, David J. and Moore, Liam and Russell, Michael and White, Chris D.",
    title = "{Constraining top quark effective theory in the LHC Run II era}",
    eprint = "1512.03360",
    archivePrefix = "arXiv",
    primaryClass = "hep-ph",
    doi = "10.1007/JHEP04(2016)015",
    journal = "JHEP",
    volume = "04",
    pages = "015",
    year = "2016"
}

@article{Jiang:2025frv,
    author = "Jiang, Xu-Hui and Liu, Yiming and Yan, Bin",
    title = "{Probing top-quark electroweak couplings indirectly at the Electron-Ion Collider}",
    eprint = "2507.21477",
    archivePrefix = "arXiv",
    primaryClass = "hep-ph",
    reportNumber = "CPTNP-2025-022",
    doi = "10.1103/w9wl-cjzq",
    journal = "Phys. Rev. D",
    volume = "112",
    number = "11",
    pages = "L111303",
    year = "2025"
}

@article{Cao:2020npb, 
    author = "Cao, Qing-Hong and Yan, Bin and Yuan, C. P. and Zhang, Ya", 
    title = "{Probing $Zt\bar{t}$ couplings using $Z$ boson polarization in $ZZ$ production at hadron colliders}", 
    eprint = "2004.02031", 
    archivePrefix = "arXiv", 
    primaryClass = "hep-ph", 
    reportNumber = "MSUHEP-20-004", 
    doi = "10.1103/PhysRevD.102.055010", 
    journal = "Phys. Rev. D", 
    volume = "102", 
    number = "5", 
    pages = "055010", 
    year = "2020" 
}

@article{Maltoni:2019aot,
    author = "Maltoni, Fabio and Mantani, Luca and Mimasu, Ken",
    title = "{Top-quark electroweak interactions at high energy}",
    eprint = "1904.05637",
    archivePrefix = "arXiv",
    primaryClass = "hep-ph",
    reportNumber = "CP3-19-16",
    doi = "10.1007/JHEP10(2019)004",
    journal = "JHEP",
    volume = "10",
    pages = "004",
    year = "2019"
}
\end{document}